\documentclass[10pt,journal,compsoc, twocolumn]{IEEEtran}
\ifCLASSOPTIONcompsoc
  \usepackage[nocompress]{cite}
\else
  \usepackage{cite}
\fi
\ifCLASSINFOpdf
\else
\fi
\usepackage{amsmath}
\usepackage{bm}
\usepackage{amsfonts}
\usepackage{caption}

\usepackage{algorithmic}
\usepackage{graphicx}
\usepackage{multirow}
\usepackage{subcaption}
\usepackage{lipsum}
\newtheorem{definition}{Definition}
\newtheorem{problem}{Problem}
\newtheorem{example}{Example}

\begin{document}

%
\title{Reinforcement Learning based Path Exploration for Sequential Explainable Recommendation}

\author{\IEEEauthorblockN{Yicong Li\IEEEauthorrefmark{1},
Hongxu Chen\IEEEauthorrefmark{1} \IEEEauthorrefmark{2},
Yile Li, 
Lin Li, \IEEEmembership{Member, IEEE,}
Philip S. Yu, \IEEEmembership{Fellow, IEEE,}
Guandong Xu\IEEEauthorrefmark{2}, \IEEEmembership{Member, IEEE} }

\IEEEcompsocitemizethanks{
\IEEEcompsocthanksitem Yicong Li, Hongxu Chen and Guandong Xu are with Data Science and Machine Intelligence Lab, Faculty of Engineering and Information Technology, University of Technology Sydney, Sydney, New South Wales, 2007, Australia.
E-mail: Yicong.Li@student.uts.edu.au, {\{Hongxu.Chen, Guandong.Xu\}}@uts.edu.au
\protect\\
\IEEEcompsocthanksitem Yile Li is with School of Transportation Engineering, Tongji University, Shanghai, 200092, China.
E-mail: 351614132@qq.com\protect\\
\IEEEcompsocthanksitem Lin Li is with College of Computer Science and Technology, 
Wuhan University of Technology, Wuhan, 430070, China.
E-mail: cathylilin@whut.edu.cn\protect\\
\IEEEcompsocthanksitem Philip S. Yu is with 
Department of Computer Science, University of Illinois at Chicago, Chicago, Illinois, 60637, the United States.
E-mail: psyu@cs.uic.edu\protect\\
}
\thanks{* Both authors contributed equally to this research.}
\thanks{\dag Corresponding author.}}

\markboth{IEEE Transactions on Knowledge and Data Engineering}
{Yicong \MakeLowercase{\textit{et al.}}: Reinforcement Learning based Path Exploration for Sequential Explainable Recommendation}

%



\IEEEtitleabstractindextext{%
\begin{abstract}
Recent advances in path-based explainable recommendation systems have attracted increasing attention thanks to the rich information provided by knowledge graphs. Most existing explainable recommendations only utilize static knowledge graphs and ignore the dynamic user-item evolutions, leading to less convincing and inaccurate explanations. Although there are some works that realize that modelling user's temporal sequential behaviour could boost the performance and explainability of the recommender systems, most of them either only focus on modelling user's sequential interactions within a path or independently and separately of the recommendation mechanism. In this paper, we propose a novel \textbf{\underline{T}}emporal \textbf{\underline{M}}eta-path Guided \textbf{\underline{E}}xplainable \textbf{\underline{R}}ecommendation leveraging \textbf{\underline{R}}einforcement \textbf{\underline{L}}earning (\textbf{TMER-RL}), which utilizes reinforcement item-item path modelling between consecutive items with attention mechanisms to sequentially model dynamic user-item evolutions on dynamic knowledge graph for explainable recommendation. Compared with existing works that use heavy recurrent neural networks to model temporal information, we propose simple but effective neural networks to capture users' historical item features and path-based context to characterize the next purchased item. Extensive evaluations of TMER on two real-world datasets show state-of-the-art performance compared against recent strong baselines. 
\end{abstract}

\begin{IEEEkeywords}
Reinforcement learning, sequential recommendation, meta-path, explanation.
\end{IEEEkeywords}}

\maketitle

\IEEEdisplaynontitleabstractindextext

%
\IEEEpeerreviewmaketitle

\section{Introduction}
Reasoning with paths over user-item associated Knowledge Graphs (KGs) has been becoming a popular means for explainable recommendations \cite{xian2019reinforcement, wang2019explainable, zhao2020leveraging}. The path-based recommendation systems have successfully achieved promising recommendation performance, as well as appealing explanations via searching the connectivity information between users and items across KGs. One category of existing works on path-based explainable recommendations seeks auxiliary meta-paths (pre-defined higher-order relational compositions between various types of entities in KGs) as similarity measures and evidence for possible explanations between users and items. 

\begin{figure}
\begin{subfigure}{.5\textwidth}
  \centering
  \includegraphics[scale=0.4]{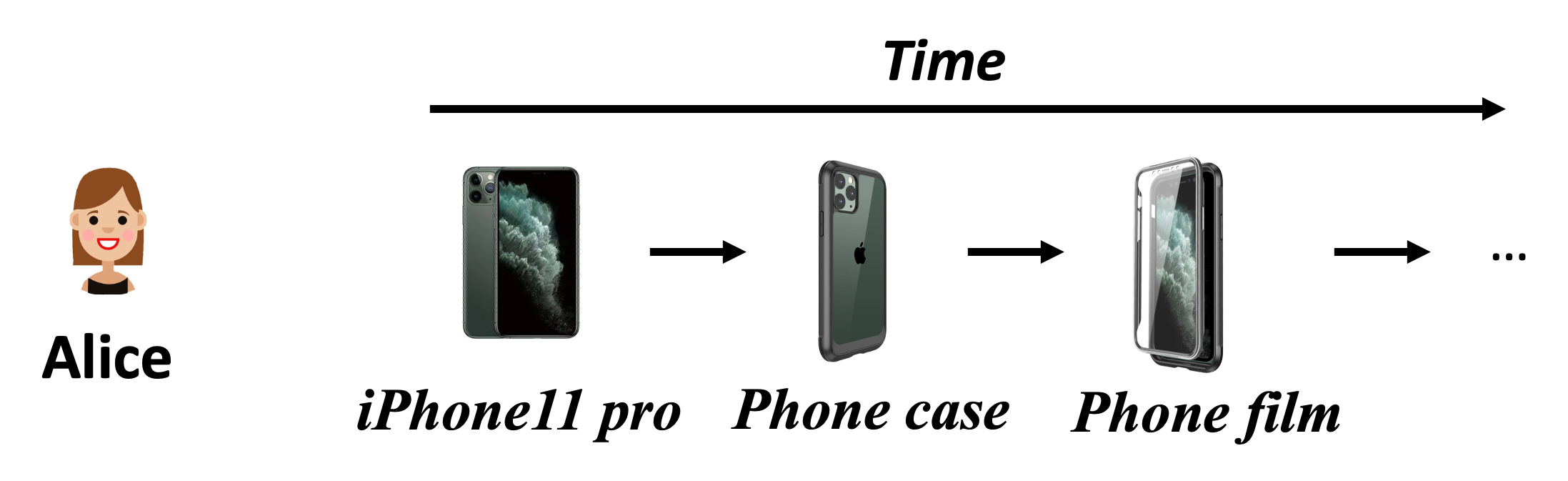}  
  \vspace{-0.5em}
  \caption{Intuitive sequential modelling}
  \label{fig:1-1}
\end{subfigure}

\vspace{1.5em}

\begin{subfigure}{.5\textwidth}
  \centering
  
  \includegraphics[scale=0.4]{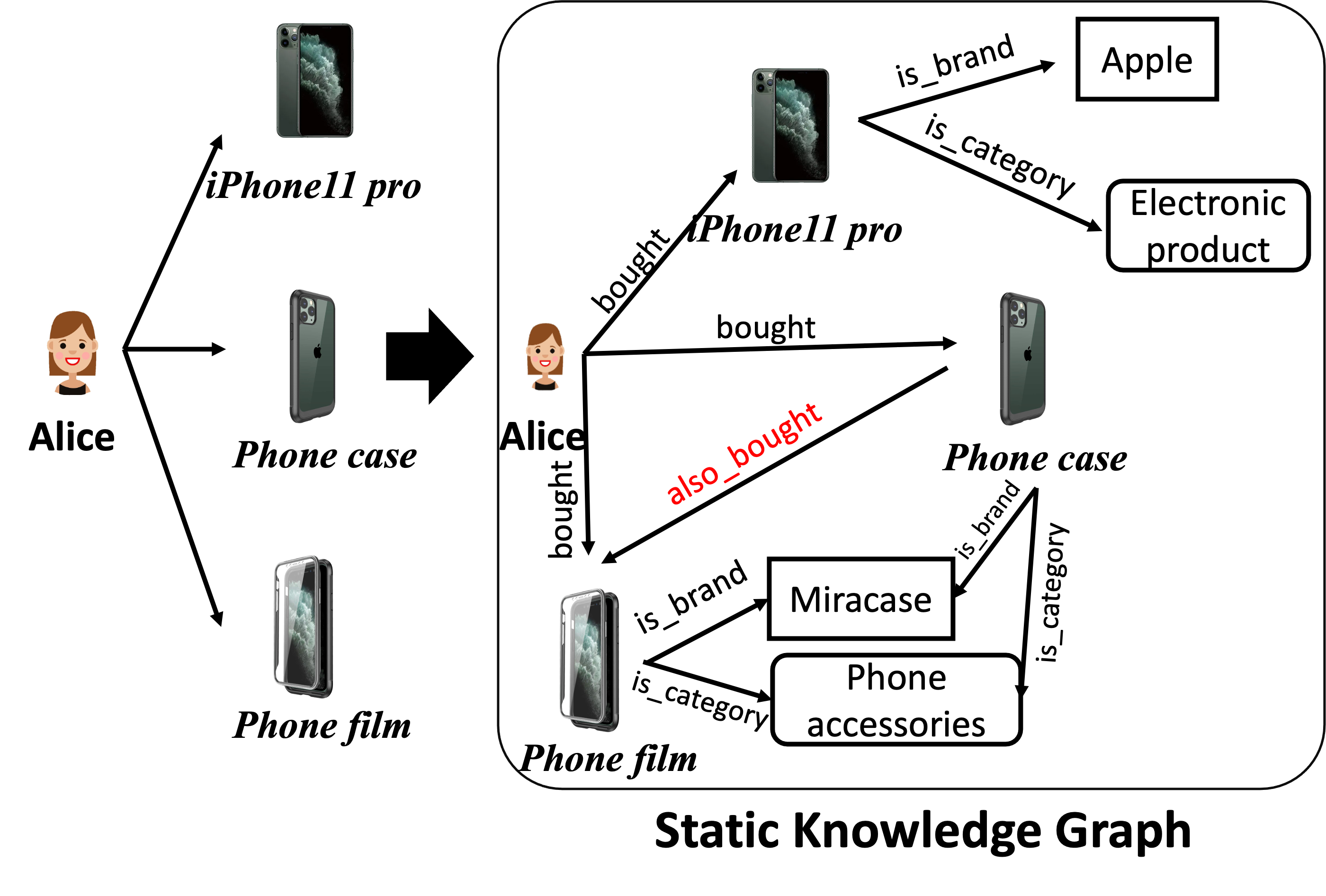}  
  \vspace{-0.5em}
  \caption{Existing static knowledge-graph based modelling}
  \label{fig:1-2}
\end{subfigure}

\caption{Existing Sequential modelling and static knowledge-graph based modelling.}
\label{fig:1}
\end{figure}

However, most of existing path-based methods for explainable recommendation simply treat the underlying KGs as static graphs, ignoring the dynamic and evolving nature of user-item interactions in real-world recommendation scenarios. The dynamics and evolutions of users' interactions with items play a pivotal role in both recommendation precision and explanations for a users' real intent. Take the scenario in Figure \ref{fig:1} as an example, Alice purchased a phone case and a phone film after a recent buy of a new phone. If we ignore the sequential information between each purchase (in Fig \ref{fig:1-1}) and treat the whole information as a static graph (in Fig \ref{fig:1-2}), the system probably gives an explanation of buying the phone case is because a similar customer also bought this phone case by exploring co-purchasing relationships. Whilst the explanation, in this case, might be valid and the underlying reasoning mechanism (collaborative filtering signal) can be used for recommendation, it is still sub-optimal as a more appealing motivation of buying a phone case is due to the recent purchase of a new phone. For this reason, in this example, it is important for a system to be capable of considering temporal and sequential user-item interactions and disentangling the importance of various reasons when generating possible explanations. 

Although some prior works have considered some extent of the sequential information for knowledge-aware path-based explainable recommendation problem, they still fail to explicitly model the dynamics of users' activities. To allow effective reasoning on paths to infer the underlying rationale of a user-item interaction, the method proposed in \cite{wang2019explainable} takes the advantages of path connectivity and leverages the sequential dependencies of entities and sophisticated relations of a path connecting a user-item pair. Nevertheless, the methods only consider the sequential information within specific paths and did not consider the importance of the user' historical sequences on reflecting the user's dynamic interactions with items. To improve, the KARN model \cite{zhu2020knowledge} fuses the user's clicked history sequence and path connectivity between users and items in KGs for recommendation. However, the method models the user's sequential behaviours and user-item interactions separately and in a coarse-grained manner (treating a user's click history as a whole), which may restrict the expressiveness of users' temporal dynamics on recommendation explainability. 

In light of this, in this paper, we challenge the problem of exploring users' temporal sequential dynamics in the context of path-based knowledge-aware recommendation. Different from existing works that either only consider the sequential information within a path or treat the user's sequential interactions as a whole and separately, we aim to 1) explicitly model and integrate the dynamic user-item interactions over time into the path-based knowledge-aware recommendation, and 2) leverage the captured dynamics of user-item interactions to improve the performance and explainability of the recommendation.

It is worth noting that modelling users' temporal sequential behaviour with path-based knowledge-aware explainable recommendation is a non-trial and challenging task. First, path-based knowledge-aware recommender systems are built upon the well-constructed KGs, and its expected accuracy and explainability are highly related to the underlying KGs and distilled paths. If temporal information between users and items is considered, the original underlying static KGs will be cast into multiple snapshots w.r.t. the timestamps. Comparing with the static KGs that consist of all users' full timelines, each snapshot at a certain timestamp only has partial observations of the global knowledge, which will result in inferior recommendation performance. To deal with the issue, we devise a general framework for temporal knowledge aware reinforcement path-based explainable recommender systems, namely \textbf{Temporal Meta-path Guided Explainable Recommendation leveraging Reinforcement Learning (TMER-RL)}. TMER-RL provides a solution that can explicitly depict users' sequential behaviour while being able to be aware of global knowledge of the entire underlying KGs. Specifically, to model the temporal dynamics between users and items, TMER-RL naturally models the task as a sequential recommendation problem and takes as input users and their corresponding sequential purchase history. To learn the sequential dependencies between consecutive items purchased by a user, TMER-RL novelly explores item-item paths between consecutive items and embed the paths as context with elaborated designed attention mechanisms to model the dynamics between user and items. Compared to existing path-based explainable recommendation systems that only consider user-item paths as evidence and support for the recommendation decision, TMER-RL contributes another creamy layer on the top of existing works, which makes use of the powerful expressiveness of temporal information between users and items.

In addition, when taking paths into consideration for the above purpose, another challenge lies in the existence of a large number of possible paths. It is time-consuming for a model to select several paths that are meaningful, expressive, and have positive impacts on both recommendation performance and explainability.
To this end, inspired by prior work \cite{hu2018leveraging}, our previous work TMER \cite{10.1145/3437963.3441762} leverages the concept of meta-path \cite{sun2011pathsim} and explore diverse meta-path schemas to characterize the context of dynamic interactions between users and items. However, it has to pre-define meta-paths and randomly sample path instances from all generated ones, which involves human and random factor in the following recommendation module. To explore paths effectively, we utilize reinforcement learning with our designed useful score function for paths to mine diverse meta-path instances. With resort to the powerful transformer model for sequential modelling, we also elaborate item-item path attention and user-item path attention units to learn combinational features of multiple paths to further characterize users' temporal purchasing motivations and their general shopping tastes, respectively. The rationale for developing such path-based attention units is that a user's motivation towards buying a certain product is complex and consists of multiple factors. For example, when buying a new phone, a customer may consider several factors including a phone's intrinsic features such as functionality, display, camera, etc., as well as other external factors such as the choices of their close friends, and their previous purchase of certain related electronics using the same operating system. With help of the designed path-based attention units, the proposed TMER-RL framework is able to learn different weights for various possible paths which will be then used as explanations for recommendations (we show the effectiveness of using the proposed path-based attention units for the explainable recommendation in the experiments by running case studies). 

Another shining point of our work is that the introduced item-item path modelling also serves as the core of the simple yet effective sequential modelling architecture of TMER-RL. The combinational meta-path enriched features learned by the item-item path attention units are aggregated with features of previous purchased item will serve as the prediction signals for the next item. The intuition behind is that the item-item paths connect two consecutive items purchased by a user, and represent various reasons and factors that may lead to the next purchased item, which intrinsically provides strong sequential dependencies between items. Compared to existing works on the sequential recommendations that rely on Recurrent Neural Nets such as (RNNs\cite{liu2016context}, GRUs\cite{hidasi2018recurrent}, LSTM\cite{wang2019explainable, zhu2020knowledge}), the proposed methods in this paper is light, simple but effective.

The main contributions of this paper are summarized as follows:
\begin{itemize}
	\item We point out that explicitly modelling dynamic user-item interactions over time can significantly benefit the recommendation performance and explainability. We introduce, to the best of our knowledge, the first study to model users' temporal sequential behaviour with path-based knowledge-aware explainable recommendation.
	
	\item We propose \textbf{T}emporal \textbf{M}eta-path Guided \textbf{E}xplainable \textbf{R}ecommendation leveraging \textbf{R}einforcement \textbf{L}earning (\textbf{TMER-RL}), which considers users' dynamic behaviours on top of the global knowledge graph for sequential-aware recommendation and explores both user-item and item-item meta-path paths with well-designed reinforcement framework and attention mechanisms for explainable recommendation.

	\item Extensive evaluations on three real-world benchmark datasets have been conducted, and the experimental results demonstrate the superiority, effectiveness and temporal explainability of the proposed TMER-RL model. 
	
\end{itemize}
\section{Problem Definition}

In this section, we first give some essential definitions and define the problem. 
\label{sec_Problem_Definition}
\begin{definition}
	\textbf{Information Networks.} An information network is a simple graph $G=(\mathcal{V},E)$. Each edge $e\in E $ represents a particular relation $r \in R$ of two entities$(v_1,v_2)$ linked by edge $e$. Each entity $v\in \mathcal{V}$ belongs to a particular type $T$. $(v_1,r,v_2)$ is a triplet means head entity $v_1$ to tail entity $v_2$ is with relation $r$. In general, relation $r$ and its reverse ${r}^{-1}$ is not the same unless the relation is symmetrical.
\end{definition}
\begin{definition}
	\label{pro_def_2.2}
	\textbf{Heterogeneous Information Network.} A heterogeneous information network (HIN) is a special type of information network. In heterogeneous information network $H$, edges $E$ and entities $\mathcal{V}$ are in different types, that is, $|R|>1, |T|>1$.
\end{definition}

\begin{figure}[ht]
	\centering
	\includegraphics[width=0.5\textwidth]{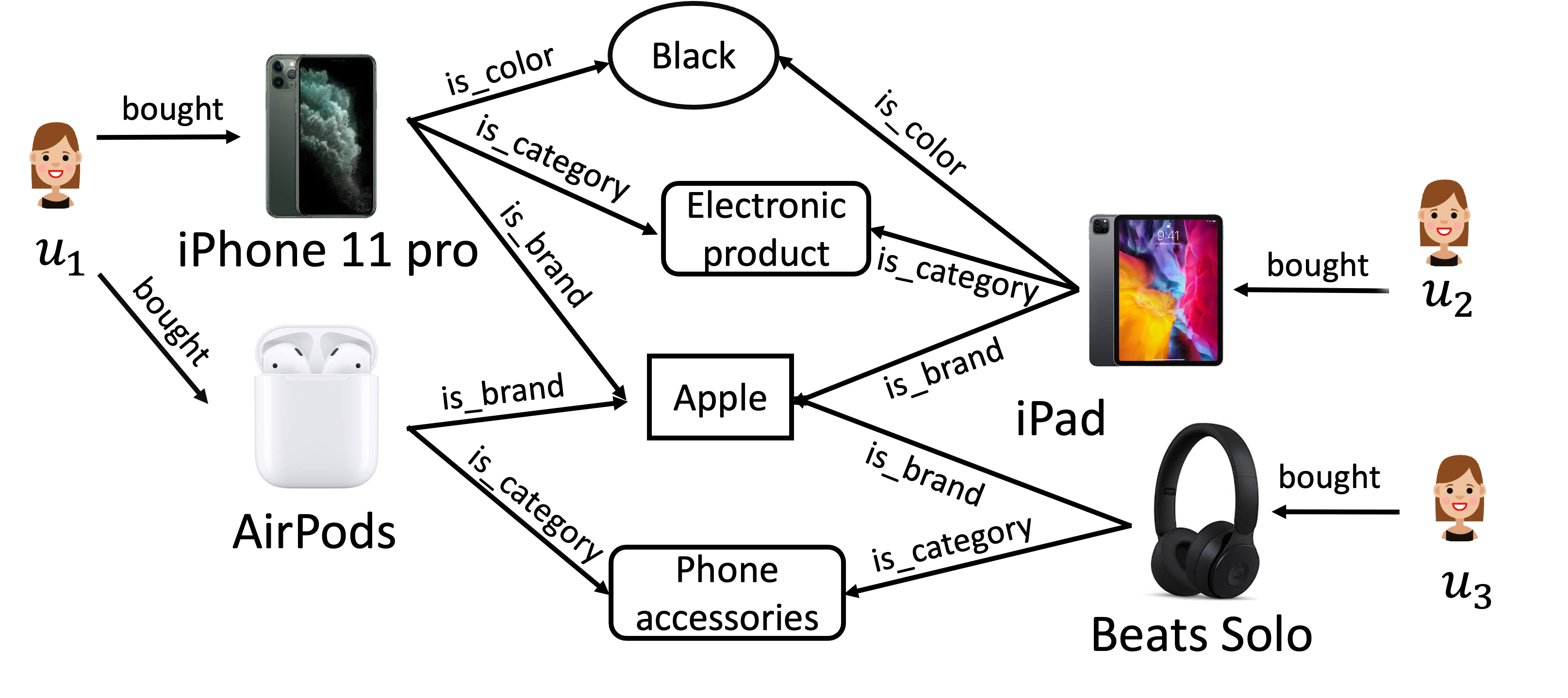}
	\caption{An example of a heterogeneous information network: product recommendation network} \label{example_product}

\end{figure}

\begin{example}
	As shown in Figure \ref{example_product}, A product recommendation network $G_P$ is a HIN, in which entities $\mathcal{V}$ could be many types, including items, users, categories, colors, brands, etc. $E$ could represent different relations. Edges between users and items denote the {\itshape buy} relation, edges between items and brands denote the {\itshape is brand of} relation, etc.
\end{example}

\begin{definition}
	\textbf{Meta-path}. A meta-path \cite{sun2011pathsim} $P$ in a network from entity $v_0$ to $v_k$, is denoted as $v_0\stackrel{r_0}{\longrightarrow}v_1\stackrel{r_1}{\longrightarrow}v_2\stackrel{r_2}{\longrightarrow}...\stackrel{r_{k-1}}{\longrightarrow}v_k$, where composite relation from $v_0$ to $v_k$ is $r=r_0 \circ r_1\circ r_2\circ ...\circ r_{k-1}$. $\circ$ represents the composition operator on relations.
\end{definition}
\begin{example}
	In the product recommendation network $G_P$ in Figure \ref{example_product}, there are many meta-paths; one of the meta-paths is $UIBI$, which means $user \ {\rightarrow} \ item \ {\rightarrow} \ brand \ {\rightarrow} \ item$. In the paper, we use the meta-path instances instead of meta-paths. A meta-path instance is a specific path, like $u_1 \ {\rightarrow} \ AirPods \ {\rightarrow} \ Apple \ {\rightarrow} \ Beats\,Solo$ is a meta-path instance of meta-path $UIBI$.
\end{example}

\begin{problem}\textbf{Sequential knowledge-aware explainable recommendation.} For a user $u_i \in U$, given the item set $I$, the user transaction sequence $\pi_u$ and the item associated information network $G$, the target of knowledge-aware explainable recommendation is to predict top $k$ items that $u_i$ will interact with, as well as the possible reasoning of recommended items. 
\end{problem}

\section{Methodology}

\begin{figure*}[h]
	\centering
	\includegraphics[width=0.8\textwidth]{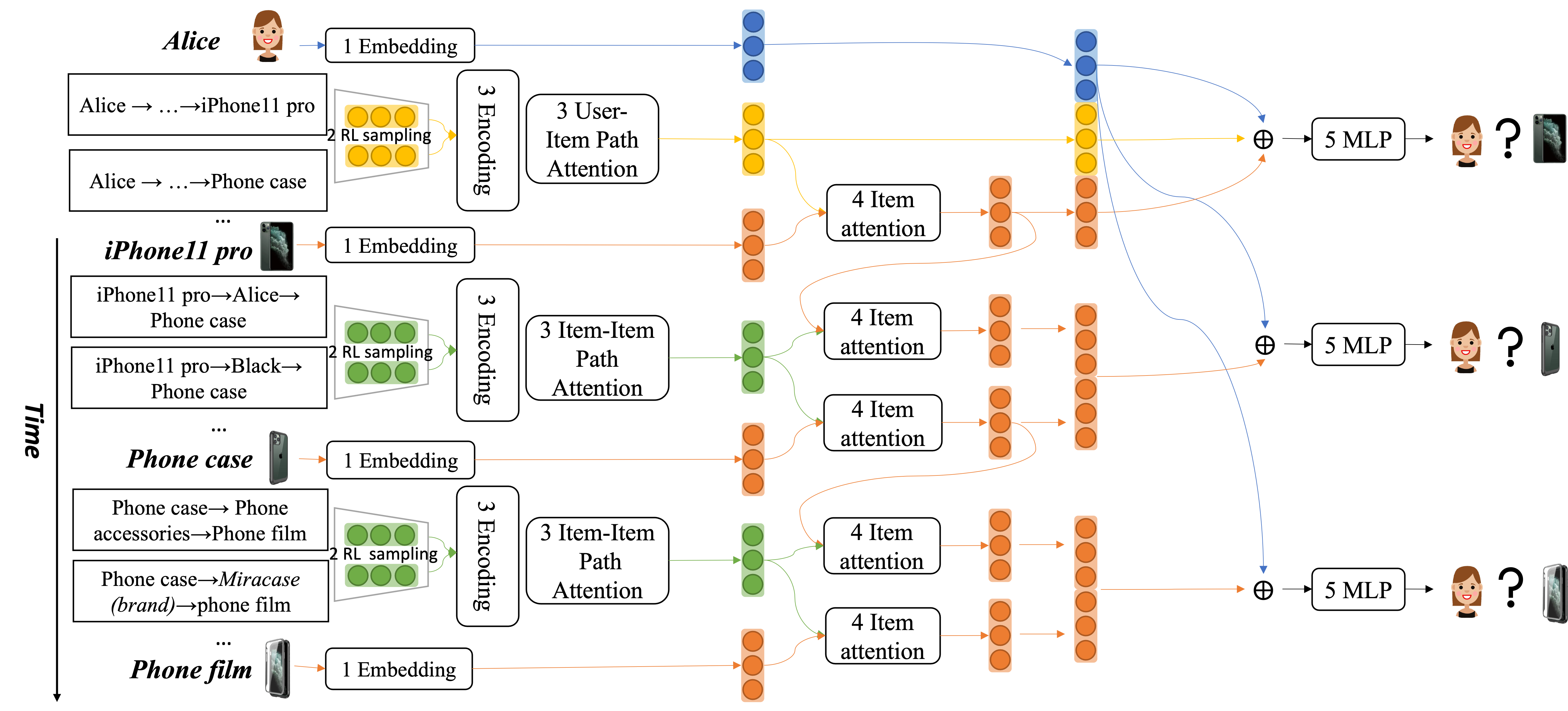}
	\caption{It is the architecture of Reinforcement Learning based Temporal Meta-path Instances Guided Explainable Recommendation (TMER-RL). Here shows an example: User \textit{Alice} bought \textit{iPhone 11 pro}, \textit{Phone case} and \textit{Phone film} in a sequential order and the training process for \textit{Alice}. The number in each network block is the step order of the model. Details are in Section \ref{sec:3.1}.} \label{architecture}
\end{figure*}
In this section, we introduce the proposed model Reinforcement Learning based Temporal Meta-path Instances Guided Explainable Recommendation (TMER-RL). In the remaining of the paper, we use the notation summarized in Table \ref{table_notations} to refer to the variables and parameters used throughout the paper.

\begin{table}[t]
	\centering
	\begin{tabular}{c|c}
		\hline
		\textbf{Symbol}  & \textbf{Description}                          \\ \hline
		$G$              & {Information network}    \\\hline
		$H$              & Heterogeneous information network (HIN)       \\\hline
		$e\in E$         & An edge                  \\\hline
		$v\in \mathcal{V}$         & An entity or node            \\\hline
		$R$            	 & Type of edges (relations) \\\hline
		$T$            	 & Type of nodes (entities)   \\\hline
		$\mathcal{U}$      & User entity set     \\ \hline
		$\mathcal{I}$      & Item entity set     \\ \hline
		$\mathcal{B}$      & Brand entity set    \\ \hline
		$\mathcal{C}$      & Category entity set \\ \hline
		$\mathcal{S}$      & State set \\ \hline
		$\mathcal{A}$      & Action set \\ \hline
		$p\in P$         & A meta-path in an HIN                              \\\hline
		$u\in U$         & User $u$                     \\\hline
		$i\in I$         & Item $i$                    \\\hline
		$W$              & Weight matrix                                 \\\hline
		$b$              & Bias vector                                   \\\hline
		$\phi \in \Phi$  & Path instance in a HIN                               \\\hline
		$\mathcal{N}$      & Neighbour function \\ \hline
		$h$              & Embedding                                     \\\hline
		$\alpha$         & Attention mechanism parameter                 \\ \hline
		$\hat r$              & User-item rating                             \\ \hline
	\end{tabular}
	\caption{\label{table_notations}Description of major notations used in this paper}
\end{table}

\subsection{Overview of TMER-RL Architecture}
\label{sec:3.1}
The overall architecture of the proposed TMER-RL model is shown in Figure \ref{architecture}. It mainly consists of five components. First, to initialize users and items, we use DeepWalk \cite{perozzi2014deepwalk} to embed user and item entities. Secondly, 
instead of pre-defined meta-paths and randomly generating meta-paths instances \cite{10.1145/3437963.3441762}, we utilize a reinforcement learning with a Markov Decision Process (MDP) environment to explore useful and meaningful sequential (temporal) and non-sequential paths to improve the recommendation performance and personalization. 
In this step, we obtain item-item instance paths between consecutive items using reinforcement learning framework. In the third step, after embedding instances, we use the multi-head attention to learn the weights of instances as the weights of reasoning paths for the specific user. Next, for the item updating, we employ a two-layer attention to make items contain reasoning information. 
This step also models the users' sequential purchased information, feeding the previous item's feature to the next one. Finally, we feed item embeddings, user embeddings, and instances to the recommendation network to do recommendation. The specific steps are elaborated in the following subsections.

\subsection{Initialize User and Item Representations}
\label{subsec3.2}
We firstly learn latent representations for involved users and items by treating random truncated random walks in a user-item bipartite network as an equivalent of sentences in DeepWalk \cite{perozzi2014deepwalk}, which optimizes the co-occurrence probability among the entities in a walk by using skipgram based on word2vec \cite{mikolov2013efficient}. Other recent advanced graph-based embedding initialization methods can be also applied, like GraphSage \cite{graphsage}, GAT \cite{velivckovic2017graph}, and so on. These recent works usually can even outperform than DeepWalk. However, through extensive comparison with these methods, DeepWalk is the best choice, the possible reason behind this is because DeepWalk pays more attention to the embedding of nodes in a path, while GCNs, such as GraphSage, GAT, etc., learn embedding of each entity with the aggregated feature information from its local neighborhood. That is, these approaches pay more attention to local relationships. However, if only considering the recommendation task itself, although the local relationships are significant, the global higher-order relationships learned by walks on a graph also play a notable role. For example, the transitivity of co-purchasing relationships between friends and the substitute relationship of items can be discovered by modelling higher-order relations in a bipartite user-item graph.

\subsection{Incorporating Meta-path based Context}
\label{subsec_3.3}
In recommendation tasks, external features related to users and items, such as product attributes, user social network, and user demographic information are usually considered as additional auxiliary information to complement traditional collaborative filtering methods. However, how to utilize the heterogenous additional information efficiently is an open problem. Some prior works \cite{jamali2010matrix, ma2008sorec, yuan2011factorization} attempted to consider social relations as the user-side information to boost the recommendation performance.
To seek help from injecting more complex additional information, recent works \cite{yu2014personalized, han2018aspect} introduced meta-paths into recommendation methods to describe relational compositions between various types of entities in heterogeneous information networks. In \cite{yu2014personalized}, the authors proposed to diffuse user preferences along different meta-paths in information networks to generate latent features of users and items. Related work \cite{han2018aspect} firstly extracts different-aspect features with meta-paths from a HIN, and then fuse aspect-level latent factors to the recommendation systems. However, these methods largely rely on the latent factors obtained from constructed meta-path based similarity matrix, which are too general and only can reflect mutual interaction between different types of entities in a graph but cannot capture the specific information along particular path instances. Therefore, inspired by existing work \cite{hu2018leveraging}, our prior work TMER and the proposed model TMER-RL explore to improve both recommendation performance and explainability by modelling more specific meta-path instances. 

Different from existing works, we differentiate meta-path instances into two different categories based on the involved entities in a recommendation scenario (i.e., user-item and item-item meta-path instances). Through modelling these path instances, we learn a more detailed meta-path based context to further characterize the motivations, reasons as well as leading factors between each pair of user-item interaction. While previous works such as \cite{hu2018leveraging} mainly focuses on modelling meta-path instances between a user and an item, this paper highlights the item-item meta-path instances, which we think is beneficial in multiple aspects to sequential explainable recommender systems. Firstly, only considering user-item paths is restrictive for recommendation explainability as user-item paths only represent a user's general shopping interests. In comparison, item-item paths are more expressive and can reflect devise reasons by exploring higher-order relations among items, such as complementary products (e.g., phone and a phone case), substitutable items of known items (e.g., iPhone - phone film - Huawei Phone), co-purchased products with other people, etc. In addition, item-item paths sometimes also serve as sequential modelling signals that naturally capture the temporal dependencies between each consecutive item purchased by users, which will be of great impact for sequential explainable recommendations.

Despite the powerful expressiveness of meta-paths in exploring HIN based knowledge-aware recommendation, it is still challenging mainly because the number of meta-paths is too large to handle (i.e., the amount of edges is cubic to the number of entities). Taking the electric product recommendation situation for an example, for $IUIBI$ meta-path schema, if we fix the starting node - iPhone11 pro, there are many instances: $iPhone11 \ pro\rightarrow Alice\rightarrow iPad \rightarrow Apple\rightarrow AirPods$, $iPhone11 \ pro\rightarrow Amy\rightarrow Phone \ case \rightarrow Miracase \rightarrow Phone \ film$, and so on. Therefore, it is necessary to explore useful path instances while limiting the total amount to simplify later calculations. The path instances exploration is introduced in the next subsection.

\subsection{Reinforcement Learning for Paths Exploration}
\label{sec:reinforcement_learning}
Our previous work \cite{10.1145/3437963.3441762} pre-defines use-item and item-item meta-paths on the recommendation knowledge graph, and randomly selects meta-path instances from all existing ones. The hand-crafted meta-paths not only need human efforts, but also are difficult to be determined when dealing with large recommendation knowledge graphs. Therefore, we propose a reinforcement learning module to explore potential \textit{useful} meta-path instances on the recommendation knowledge graph. 
The definition of \textit{usefulness} in our work is to have contribution to the recommendation module learning. The process of meta-path instances exploration can be defined as a Markov Decision Process (MDP)\cite{shani2005mdp}, and we use reinforcement learning to train the exploration policy. 

\begin{itemize}
	\item \textit{State.}
	It is the status of each step and the state set is defined as $\mathcal{S}$. At step $t$, the state $\mathcal{S}_t$ is $(e_t, his_t)$, where $e_t \in (\mathcal{U} \cup \mathcal{I} \cup \mathcal{B} \cup \mathcal{C})$ is the current visiting entity and $his_t = \{e_0, e_1, ..., e_t\}$ is the history visited entities including the current one. Inspired by \cite{xian2019reinforcement, saebi2020heterogeneous, wenhan_emnlp2017, minerva}, we add self-loop and inverse edges on knowledge graph $G$ to facilitate graph traversal. 
	\item \textit{Action.} 
	The action set $\mathcal{A}_t$ at step $t$ means all candidate entities to go. We define $\mathcal{A}_t = \mathcal{N}(e_t) \cup e_t$, which means the the candidate entities are the neighbours of the current visiting entity $e_t$ and itself. Moreover, we allocate each action a weight to present the probability of choosing each action. The score function of each action at step $t$ is as follows.
	\begin{equation}
		s_{e_i} = softmax(\frac{h_{e_t} \cdot h_{e_i}^T}{||h_{e_t}|| ||h_{e_i}||}),\  where\ e_i \in \mathcal{A}_t
	\end{equation}
	where $h_{e_t}$ and $h_{e_i}$ mean the presentation of current visiting entity and one of the actions, respectively. We believe cosine function could indicate the similarity of two entities and it is more likely to choose similar entities to form a path on the knowledge graph. However, on the knowledge graph, some entities have dense structure, leading to large action set. Thus, pruning unimportant candidate entities is necessary to improve efficient. We rank each actions score descendingly and obtain top $k$ as new action set.
	
	\item \textit{Transition.}
	Given a current state $\mathcal{S}_t=(e_t, his_t)$ and an action $e_i \in \mathcal{A}_t$, the transition probability to the next state $\mathcal{S}_{t+1}=(e_{t+1}, his_{t+1})$ is 
	\begin{equation}
		p(\mathcal{S}_{t+1}|\mathcal{S}_t,e_i)=1
	\end{equation}
	\item \textit{Reward.}
	The target of the reinforcement learning is to find user-item meta-path instances and item-item meta-path instances, so the target entity is known for each user. Therefore, if get to the target entity before pre-defined maximum step $T$, the reward is 1, otherwise 0, mathematically,
	\begin{equation}
		\mathcal{R}=
		\left\{
		\begin{array}{lr}
			1, \ if\ get\ to\ the\ target\ entity\ before\ step\ $T$,&  \\
			0, \ otherwise &  
		\end{array}
		\right.
	\end{equation}
	
	\item \textit{Optimization.}
	The whole reinforcement learning module maximizes the final reward to learn policy $\pi$ to find useful user-item and item-item meta-path instances, mathematically,
	\begin{equation}
		J(\theta) = \mathbb{E}_{e_0\in (\mathcal{U} \cup \mathcal{I})} (\mathbb{E}_{\pi}(\mathcal{R}))
	\end{equation}
	where $e_0$ is the initial entity. We use REINFORCE \cite{williams1992simple} algorithm to train the objective function.
\end{itemize}

According to the above MDP framework, we can get candidate user-item and item-item meta-path instances. Taking user-item meta-path instances from $u_n$ to $i_m$ for an example, there should be $p$ candidate paths. We use a score function to calculate each path's score, mathematically,
\begin{equation}
	c_{u_n \rightarrow i_m} = \frac{\sum_{t=1}^{T} s_{e_t}}{T}
\end{equation}
where $s_{e_t}$ is each step's action score. The rationale to choose average step's score as the path's score is that we do not want the length of path to influence the path's score. We rank the score of paths descendingly and get top $q$ candidate paths. 

\subsection{Parameterizing Combinational Features of Meta-paths as Recommendation Context}

\label{subsec_seq_and_non_seq}

\subsubsection{\textbf{Learning combinational path-based features with Self-Attention}}
\label{subsubsec3.3.1}
After obtaining candidate user-item meta-path instances and item-item meta-path instances, we first regard paths as sentences, nodes as tokens in sentences, using Word2Vec\cite{mikolov2013efficient} method and $Mean(\cdot)$ operations to learn path embedding. Then, we employ multi-head self-attention units to learn the meta-path based context (the User-Item and Item-Item Path Attention modules shown in Figure \ref{architecture}). The rationale of deploying such self-attention units here is that because, after sampling, there are still multiple paths between each pair of item-item (or user-item) representing particular distinct reasons (reasoning paths); and we observe that the reasons for buying consecutive two items are not simply unique; rather, the reasons are more complex and likely a mixture of multiple different reasons. For example, the reasons for a customer to buy a phone case right after his/her previous purchase of a new phone are probably a mixture of 1) his/her friends who own a similar phone and bought this particular phone case, 2) the phone case is the most popular match for the purchased new phone, 3) the color of the phone case matches the customer's preference. The potential reasons can be more than the listed, and again they can be represented by using various meta-paths.

Based on this observation, self-attentive properties of the Transformer model \cite{vaswani2017attention}, we aim to learn the combinational features from multiple path instances to better characterize the complex reasons between each connected pair of entities in the KG. 
\begin{equation}
	\begin{split}
		&Attention (Q_{\phi},K_{\phi},V_{\phi})=f(\frac{Q_{\phi}K_{\phi}^T}{\sqrt{d_k}})V_{\phi}, \\
		&MultiHead (Q_{\phi},K_{\phi},V_{\phi})=Concat (head_1,...,head_m)W^O , \\
	\end{split}
\end{equation}
where $ head_i $ is $ Attention (W_i^QQ_{\phi},W_i^KK_{\phi},W_i^VV_{\phi}). $ $f(.)$ is a softmax function. Query $Q$, key $K$ and value $V$ are self-attention variables associated with path $\phi$, and $W$ is the weight. $d_k$ is the dimensionality (here $d_k = 100$). $Concat(.)$ is the concatenation operation.

\subsubsection{\textbf{Modelling temporal dependencies with item-item meta-path instances}}
\label{sec:subsubsec3.4.2}
To learn the temporal dynamics of each user, the proposed TMER-RL framework artfully resorts on the above-mentioned item-item meta-path instances together with the well-designed architecture to capture the sequential dependencies between two consecutive items. Compared with most existing works on sequential recommendation \cite{wang2019explainable, zhu2020knowledge, xu2019recurrent} that utilize recurrent neural networks to encode the temporal effects between items in a user's interacted sequence, the proposed TMER-RL bypasses the de-facto default deployment of RNNs or LSTMs that sometimes make the model even heavier. Specifically, the proposed framework novelly model the temporal dependencies between two items by capturing 1) the information passed from the previous item though an item-attention unit, 2) item-item connectivity trough a specific candidate path instance. Notably, the information passed from the previous item is an attentive aggregation of previous item-item connectivity information. For example, in Figure \ref{architecture}, whether Alice will buy the phone film is modelled by considering 1) the information passed from the phone case (which includes the connectivity between iphone11 pro and the phone case), and 2) the paths between the phone case to the phone film. As a result, the long-range and short-term "memory" in a sequence can be captured, and the extent of the goodness of the long and short term memory can be influenced by the length of the modelled sequence in different scenarios with different datasets.

\subsubsection{\textbf{Updating item representations}}
After updating representations of user-item meta-path instances and item-item meta-path instances according to a multi-head self-attention mechanism, we employ a different kind of attention mechanism to update item representations. It is obvious that the current item is mostly related to the last one, which means it is better to add the last item's information to the current one to contain temporal information. Besides, the current item is also related to the instances that arrived at this item. Therefore, we perform a two-layer attention mechanism to update item representations. Mathematically,

\begin{equation}
	\label{eq3.4.2.2}
	h_i^{(1)}=g(W_{i-1} h_{i-1} + W_{\phi _{i-1\rightarrow i}} h_{\phi _{i-1\rightarrow i}}+b_i^{(1)}) \odot h_{i-1}
\end{equation}
\begin{equation}
	\label{eq3.4.2.3}
	h_i^{(2)}=g(W_i h_i+W_{\phi _{i-1\rightarrow i}}^{(2)} h_{\phi _{i-1\rightarrow i}}+b_i^{(2)}) \odot h_i
\end{equation}
where $h_i^{(1)}$ and $h_i^{(2)}$ mean the first and second layer output of the item attention module, respectively. $h_{i-1}$ is the last item's latent representation. $\phi _{i-1\rightarrow i}$ is the instance from the $(i-1)^{th}$ item to the $i^{th}$ item. $W$ and $b$ with different superscripts denote different variables' weights and bias. $g(.)$ is ReLU function. However, until now, here is a problem with the calculation of the first item, because there is no item before it. To solve this problem, we involve the user-item instance from user $u$ to the first item into the update of the first item, as Eq. \ref{eq3.4.2.4} shows. Actually, the instance from $u$ to the first item is really important in the recommendation, because it is the first item user has bought and most of the other bought items have a sequential relation with the first item to some extent. That is why we embed it into the first item. 
\begin{equation}
	\label{eq3.4.2.4}
	h_{i=1} = g(W_{i} h_{i}+W_{\phi_{u\rightarrow i}}h_{\phi_{u\rightarrow i}}+b_{i}) \odot h_{i}
\end{equation}
where $\phi _{u\rightarrow i}$ represents the path from user $u$ to the first item.

\subsection{The Complete Recommendation Model} 
\label{subsec3.5}

Finally, we concatenate user, item and instances information (calculated in \ref{sec:subsubsec3.4.2}) to a vector according to Eq. \ref{eq3.5.1}, and get user-item prediction scores through Multilayer Perceptron (MLP) with explainability instances.

\begin{equation}
	\label{eq3.5.1}
	h_{u,i}=[h_u;h_i^{(1)};h_i^{(2)}]
\end{equation}
where $[;]$ means vector concatenation. Here $h_{u,i}$ denotes the explicit mutual vector of the user, item, and implicit mutual of user-item meta-path instances and item-item meta-path instances. For the first item, the concatenation operation is different because of the dimension problem, and therefore, for the first item related to each user, the vector fed in MLP involves user-item instance, mathematically,
\begin{equation}
	h_{u,i=1} = [h_u;h_{\phi _{u\rightarrow 1}};h_{i=1}]
\end{equation}
After that, final user-item rating calculates as follows.
\begin{equation}
	\label{eq3.5.2}
	r_{u,i}=MLP(h_{u,i})
\end{equation}
where the MLP contains a two-hidden-layer neural network with ReLU as the activation function and sigmoid function as the output layer. According to \cite{hu2018leveraging, he2016deep}, neural network models can learn more abstractive features of data via using a small number of hidden units for higher layers, we employ a tower structure for the MLP component, halving the layer size for each successive higher layer.

We use implicit feedback loss with negative sampling \cite{he2017neural, tang2015line} as the whole loss function:

\begin{equation}
	\label{eq3.5.3}
	loss_{u, i}=-E_{j \sim P_{n e g}}\left[\log \left(1-r_{u, j}\right)\right]
\end{equation}
where models the negative feedback drawn from the noise distribution $P_{n e g}$, which is a uniform distribution following \cite{hu2018leveraging}.

\subsection{Discussion of Explanation for Recommendation with Attention Mechanism}
Attention Model \cite{bahdanau2015neural, chaudhari2019attentive} was first introduced in machine translation task and the attention weights were later widely used in natural language processing tasks as explanations in neural networks \cite{xu2015show, mullenbach2018explainable}. Other than Natural Language Processing (NLP) tasks, the attention mechanism is also a near-ubiquitous method in recommendation tasks used as explanations in some works. \cite{wang2020attention, chen2018attention, tal2019neural, cong2019hierarchical}
However, there are different opinions on whether attention mechanism could be used as a way to explain data \cite{jain2019attention, wiegreffe2019attention, serrano2019attention}.

In our proposed method TMER-RL, the item-item meta-path instances with attention weights learned by Item-Item Path Attention module in Figure. \ref{architecture} are used as explanations. To be specific, for all existing item-item meta-path instances, we use reinforcement learning to explore some useful paths, because it is difficult to process all paths whose number is large. After obtaining the candidate item-item meta-path instances, we consider these paths as the explanations for purchasing the target item. For example, in Figure. \ref{architecture}, user \textit{Alice} has bought \textit{iPhone 11 pro}, \textit{Phone case} and \textit{Phone film} in a sequential order. For paths from \textit{Phone case} to \textit{Phone film}, the paths includes $Phone \ case\rightarrow Phone \ accessories\rightarrow Phone \ film$, $Phone \ case\rightarrow Miracase\rightarrow Phone \ film$ and $Phone \ case\rightarrow Abby\rightarrow Phone \ film$. In this situation, the explanations for buying \textit{Phone film} are that \textit{Alice} has bought other \textit{Phone accessories}, \textit{Alice} has bought other \textit{Miracase} product, and that \textit{Abby} has bought the \textit{Phone case} and the \textit{Phone film} together. Based on them, we use a self-attention module as Item-Item Path Attention module to learn a distribution of these paths (explanations) to further explain the recommendation. The learned weights are considered as the explainable weighted scores for the candidate item-item instances. Detailed case study is in subsection \ref{exp:case_study}.


\section{Experiment}

In this section, we present our experimental settings and give analysis on the evaluation results.

\subsection{Experiment Settings}
\subsubsection{Datasets}

We use two public data collections to conduct experiments, Amazon datasets \footnote{http://jmcauley.ucsd.edu/data/amazon/} \cite{ni2019justifying} and Goodreads dataset \footnote{http://goodreads.com} \cite{DBLP:conf/recsys/WanM18, DBLP:conf/acl/WanMNM19}. The Amazon dataset contains 29 categories, from which we choose musical instruments dataset, automotive dataset, and toys and games dataset. Each dataset includes brand and category as the metadata. The Goodreads dataset is a public book online rating and review collection. We select mystery thriller crime genre books and choose authors and publishers as the metadata of books.
More details are shown in Table \ref{Dataset_Information}.

We select the latest twelve items for each user and order these items by timestamps, and then we choose the first two items as bridge items, the next four items as training items and the rest as test items. We create the Heterogeneous Information Networks using \textit{user}, \textit{item}, \textit{category} and \textit{brand} in Amazon datasets, and using \textit{user}, \textit{book}, \textit{author}, \textit{publisher} in Goodreads dataset, respectively. Last, user-item meta-path instances and item-item meta-path instances are explored according to section \ref{sec:reinforcement_learning}.

\begin{table}[]\centering
	
	\begin{tabular}{ccccccc}
		\hline
		\textbf{Datasets}& \textbf{User}& \textbf{Item} & \textbf{Category} & \textbf{Brand} \\ \hline
		Musical Instruments &1450&11457&429&1185\\ 
		Automotive &4600&36663&1592&3790\\ 
		Toys and Games&9300&58743&820&5404\\ 
		\hline \hline
		& \textbf{User}& \textbf{Book} & \textbf{Author} & \textbf{Publisher} \\ \hline
		Goodreads & 11800 & 12142 & 3633 & 1442\\ \hline
	\end{tabular}

	\caption{\label{Dataset_Information}Dataset Information.}
\end{table}

\subsubsection{Evaluations}

We leverage Top $K$ Hit Ratio (HR@$K$) and Top $K$ Normalized Discounted Cumulative Gain (NDCG@$K$) as our metrics. For each positive test instance ($u, i^+$), we mix the ground truth item $i^{+}$ with $\mathcal{N}$ random negative items, then rank all these $\mathcal{N}+1$ items and compute the HR@$K$ and NDCG@$K$ scores. We set $\mathcal{N}=500$ and $K=1,5,10,20$ for evaluations. To alleviate the high-bias and low-variance of sampled metrics, we use corrected evaluation metrics introduced in \cite{krichene2020sampled}.
For evaluation of the explainability of the recommendation, we use case studies to show the explanations in detail.

\subsubsection{Baselines}
\begin{itemize}
	\item \textbf{GRU4Rec} \cite{hidasi2015session, hidasi2018recurrent}: GRU4Rec is a session-based recommendation method using GRU. For each user, we treat the training items as a session. 
	
	\item \textbf{NARRE} \cite{chen2018neural}: NARRE utilizes neural attention mechanism to build an explainability recommendation system. 
	
	\item \textbf{MCRec} \cite{hu2018leveraging}: MCRec develops a deep neural network with the co-attention mechanism to learn rich meta-path based context information for recommendations. 
	
	
	\item \textbf{NFM} \cite{he2017neural2}: NFM effectively combines the linear Factorization Machines (FM) and nonlinear neural networks for prediction under sparse settings. 
	
	
	
	\item \textbf{Chorus} \cite{wang2020make}: Chorus considers both relations among items and corresponding temporal dynamics to model the recommendation data in a knowledge-aware and time-aware way. The enhanced item representations improve the performance.

	\item $\mathbf{S^3Rec}$ \cite{zhou2020s3}: $\mathrm{S^3Rec}$ utilizes the intrinsic data correlation to employ self-supervised learning tasks to learn data representations for sequential recommendation enhancement.

\end{itemize}

\subsection{Implementation Details}
\label{subsec:imple_details}
For GRU4Rec, we use ReChorus package \cite{wang2020make} to implement the algorithm. For others, we directly run the provided code by each paper. To fairly compare the evaluation results, we modify the each baseline's evaluation code as the same as TMER-RL. Especially, NARRE is a rating prediction model, we turn it into a link prediction model by rating 1 positive item and 500 negative items and ranking them. 

\subsection{Parameter Settings}
\label{subsec:param_settings}
We choose the first and second items for each user as the bridge items, the training or testing process, the last 6 items for each user as the testing positive items, and the remaining 4 are training items. 
The hyperparameters are carefully tuned to achieve optimal results in all experiments. We implement GRU4Rec, NARRE and NFM based on the paper by pytorch package. The meta-paths and settings in MCRec are the same as the original paper. The meta-paths in FMG are \textit{User-Item(UI), User-Item-Brand-Item(UIBI)} and \textit{User-Item-Category-Item(UICI)}.

In terms of the TMER compared in the following experiments, which removes the path generation via reinforcement learning module, we use \textit{UIBI}, \textit{UICI}, \textit{UIBICI} and \textit{UICIBI} as user-item meta-paths and \textit{ICIBI}, \textit{IBICI}, \textit{ICICI}, \textit{IBIBI}, \textit{IUIUI}, \textit{ICIUI} and \textit{IBIUI} as item-item meta-paths, where U, I, B and C denote user, item, brand and category, respectively. 
For our proposed TMER-RL, the learning rate for Amazon Musical Instruments dataset is $1e-5$, for the Amazon Automotive dataset is $5e-5$ and for Amazon Toys and Games is $1e-4$. The parameters for other baselines are searched for their best results. We set the maximum length of path instance exploration as 6 because we set 6 as the max length of meta-path in our previous work \cite{10.1145/3437963.3441762} and it is convenient to compare the performance.

\subsection{Effectiveness Analysis on Recommendation Results}
\label{subsec:baseline}
We compare TMER-RL with 6 other popular and recent baselines, including four sequential recommendations (GRU4Rec, Chorus,  $\mathrm{S^3Rec}$ and our former proposed TMER), an explainable recommendation (NARRE),a path-based recommendation (MCRec) and a factorization machines-based recommendation (NFM) on 4 datasets with 500 negative sampling. 

As shown in Table \ref{tab:baseline}, our proposed TMER-RL achieves the best results and in the most situations, TMER-RL could always give the correct items among 500 negative items, especially in Amazon Automotive dataset. The advantages also hold on other datasets. These results demonstrate that our framework can achieve obvious advantages through explicitly modelling each user’s sequential purchased information with the temporal paths, while NARRE and NFM ignore the sequential information for each user. MCRec ignores the item-item intrinsic relations and just utilizes user-item interactions to train the recommendation. GRU4Rec utilizes RNNs to model session-based recommendation. Chorus focuses on the way to better learning items' representations by combining the complement and substitute relations among items on the temporal items. $\mathrm{S^3Rec}$ uses different ways to pre-train embeddings to improve recommendation. However, the above three sequential recommendations all overlook the relation along paths and only focus on the temporal historical sequence, leading to worse results.

\begin{table*}[]
\centering
\begin{tabular}{|c|c|lllllllc|}
\hline
Datasets                                                                                & Metrics & \multicolumn{1}{c}{GRU4Rec} & \multicolumn{1}{c}{NARRE} & \multicolumn{1}{c}{MCRec} & \multicolumn{1}{c}{NFM} & \multicolumn{1}{c}{Chorus} & \multicolumn{1}{c}{S3Rec} & \multicolumn{1}{c}{TMER} & TMER-RL         \\ \hline
\multirow{8}{*}{\begin{tabular}[c]{@{}c@{}}Amazon \\ Music \\ Instruments\end{tabular}} & HR@1    & 0.3379                      & 0.6171                    & 0.5838                    & 0.4457                  & 0.1020                     & 0.3448                    & 0.8620                   & \textbf{0.8762} \\
                                                                                        & HR@5    & 0.6833                      & 0.8215                    & 0.6178                    & 0.5091                  & 0.5386                     & 0.5483                    & 0.9473                   & \textbf{0.9559} \\
                                                                                        & HR@10   & 0.8244                      & 0.8828                    & 0.6444                    & 0.5505                  & 0.8190                     & 0.6448                    & 0.9642                   & \textbf{0.9761} \\
                                                                                        & HR@20   & 0.9272                      & 0.9495                    & 0.6819                    & 0.6169                  & 0.9651                     & 0.7559                    & 0.9736                   & \textbf{0.9880} \\ \cline{2-2}
                                                                                        & NDCG@1  & 0.3379                      & 0.6171                    & 0.5838                    & 0.4457                  & 0.1020                     & 0.3448                    & 0.8620                   & \textbf{0.8762} \\
                                                                                        & NDCG@5  & 0.5187                      & 0.7272                    & 0.6009                    & 0.4898                  & 0.3185                     & 0.4505                    & 0.9306                   & \textbf{0.9368} \\
                                                                                        & NDCG@10 & 0.5647                      & 0.7472                    & 0.6095                    & 0.5037                  & 0.4097                     & 0.4817                    & 0.9364                   & \textbf{0.9438} \\
                                                                                        & NDCG@20 & 0.5910                      & 0.7580                    & 0.6189                    & 0.5208                  & 0.4474                     & 0.5095                    & 0.9389                   & \textbf{0.9470} \\ \hline
\multirow{8}{*}{\begin{tabular}[c]{@{}c@{}}Amazon \\ Automotive\end{tabular}}           & HR@1    & 0.4376                      & 0.6521                    & 0.6899                    & 0.6033                  & 0.1389                     & 0.4457                    & 0.9070                   & \textbf{0.9258} \\
                                                                                        & HR@5    & 0.7708                      & 0.8229                    & 0.7295                    & 0.6542                  & 0.5746                     & 0.6843                    & 0.9632                   & \textbf{0.9832} \\
                                                                                        & HR@10   & 0.8813                      & 0.8703                    & 0.7546                    & 0.6937                  & 0.8205                     & 0.7722                    & 0.9701                   & \textbf{0.9910} \\
                                                                                        & HR@20   & 0.9526                      & 0.9080                    & 0.7867                    & 0.7430                  & 0.9604                     & 0.8511                    & 0.9745                   & \textbf{0.9954} \\ \cline{2-2}
                                                                                        & NDCG@1  & 0.4376                      & 0.6521                    & 0.6899                    & 0.6033                  & 0.1389                     & 0.4457                    & 0.9070                   & \textbf{0.9258} \\
                                                                                        & NDCG@5  & 0.6146                      & 0.7450                    & 0.7100                    & 0.6383                  & 0.3587                     & 0.5737                    & 0.9550                   & \textbf{0.9734} \\
                                                                                        & NDCG@10 & 0.6506                      & 0.7604                    & 0.7180                    & 0.6517                  & 0.4386                     & 0.6019                    & 0.9574                   & \textbf{0.9761} \\
                                                                                        & NDCG@20 & 0.6688                      & 0.7699                    & 0.7261                    & 0.6644                  & 0.4746                     & 0.6221                    & 0.9585                   & \textbf{0.9773} \\ \hline
\multirow{8}{*}{\begin{tabular}[c]{@{}c@{}}Amazon\\ Toys and Games\end{tabular}}        & HR@1    & 0.3914                      & 0.6698                    & 0.6802                    & 0.5365                  & 0.1753                     & 0.4741                    & 0.8327                   & \textbf{0.8632} \\
                                                                                        & HR@5    & 0.7385                      & 0.8745                    & 0.7181                    & 0.5963                  & 0.5734                     & 0.7324                    & 0.9478                   & \textbf{0.9647} \\
                                                                                        & HR@10   & 0.8618                      & 0.9234                    & 0.7412                    & 0.6412                  & 0.7886                     & 0.8172                    & 0.9605                   & \textbf{0.9811} \\
                                                                                        & HR@20   & 0.9431                      & 0.9570                    & 0.7661                    & 0.6951                  & 0.9368                     & 0.8870                    & 0.9680                   & \textbf{0.9908} \\ \cline{2-2}
                                                                                        & NDCG@1  & 0.3914                      & 0.6698                    & 0.6802                    & 0.5365                  & 0.1753                     & 0.4741                    & 0.8327                   & \textbf{0.8632} \\
                                                                                        & NDCG@5  & 0.5745                      & 0.7813                    & 0.6996                    & 0.5775                  & 0.3773                     & 0.6117                    & 0.9292                   & \textbf{0.9456} \\
                                                                                        & NDCG@10 & 0.6147                      & 0.7973                    & 0.7070                    & 0.5929                  & 0.4472                     & 0.6400                    & 0.9337                   & \textbf{0.9512} \\
                                                                                        & NDCG@20 & 0.6355                      & 0.8059                    & 0.7133                    & 0.6069                  & 0.4852                     & 0.6578                    & 0.9356                   & \textbf{0.9534} \\ \hline
\multirow{8}{*}{Goodreads}                                                              & HR@1    & 0.2168                      & 0.5007                    & 0.4679                    & 0.0789                  & 0.1490                     & 0.1967                    & 0.5991                   & \textbf{0.6276} \\
                                                                                        & HR@5    & 0.5477                      & 0.7766                    & 0.5469                    & 0.1751                  & 0.5224                     & 0.4381                    & 0.8499                   & \textbf{0.8640} \\
                                                                                        & HR@10   & 0.7014                      & 0.8570                    & 0.5978                    & 0.2414                  & 0.7419                     & 0.5527                    & 0.9055                   & \textbf{0.9168} \\
                                                                                        & HR@20   & 0.8502                      & 0.9136                    & 0.6600                    & 0.3350                  & 0.9075                     & 0.6703                    & 0.9395                   & \textbf{0.9569} \\ \cline{2-2}
                                                                                        & NDCG@1  & 0.2168                      & 0.5007                    & 0.4679                    & 0.0789                  & 0.1490                     & 0.1967                    & 0.5991                   & \textbf{0.6276} \\
                                                                                        & NDCG@5  & 0.3910                      & 0.6494                    & 0.5083                    & 0.1451                  & 0.3379                     & 0.3232                    & 0.7968                   & \textbf{0.8116} \\
                                                                                        & NDCG@10 & 0.4441                      & 0.6755                    & 0.5247                    & 0.1677                  & 0.4091                     & 0.3603                    & 0.8162                   & \textbf{0.8305} \\
                                                                                        & NDCG@20 & 0.4818                      & 0.6899                    & 0.5404                    & 0.1919                  & 0.4514                     & 0.3901                    & 0.8251                   & \textbf{0.8395} \\ \hline
\end{tabular}\caption{Performance Comparison with Baselines.}\label{tab:baseline}
\end{table*}

\subsection{Effectiveness Analysis on Modules of TMER-RL}
\label{subsec:abtest}

To study the impact of the modules of TMER-RL for the recommendation performance, including the reinforcement learning path generation module, user-item and item-item instances modules, we future compare our model (TMER-RL) with three variants, namely TMER, $\neg$ UI and $\neg$ II. TMER removes the reinforcement learning module and generates user-item and item-item instances according to pre-defined meta-paths as mentioned in \ref{subsec:param_settings}. $\neg$ UI means we do not consider user-item instance and remove the corresponding self-attention module. $\neg$ II means we remove the self-attention module for item-item instances. We report the compared results with 100 negative sampling in Table \ref{tab:abtest}. 

TMER-RL achieves the best performance on all evaluation methods, especially on Amazon Automotive dataset.
We can see that the user-item and item-item instances self-attention modules significantly boost the recommendation method, adaptively adjust the importance of different instance paths. The self-learned user-item and item-item instance paths contribute to improving the performance of recommendation finally. Especially, using the item-item meta-path instances self-attention module helps TMER-RL improve 5.87\% and 6.85\% on HR@1 on Musical Instruments dataset and Automotive dataset, respectively.


\begin{table}[]
\begin{tabular}{c|c|c|c|c|c}
\hline
Dataset           & Metrics & TMER-RL         & TMER   & $\neg$ UI & $\neg$ II \\ \hline
\multirow{8}{*}{\begin{tabular}[c]{@{}c@{}}Amazon \\ Musical \\ Instruments\end{tabular}} & $H@1$    & \textbf{0.9252} & 0.9216 & 0.8797  & 0.8739  \\
                                            & $H@5$    & \textbf{0.9862} & 0.9855 & 0.9301  & 0.9300  \\
                                            & $H@10$   & \textbf{0.9942} & 0.9934 & 0.9366  & 0.9372  \\
                                            & $H@20$   & \textbf{0.9979} & 0.9965 & 0.9479  & 0.9471  \\ \cline{2-6} 
                                            & $N@1$  & \textbf{0.9252} & 0.9216 & 0.8797  & 0.8739  \\
                                            & $N@5$  & \textbf{0.9741} & 0.9736 & 0.9259  & 0.9247  \\
                                            & $N@10$ & \textbf{0.9765} & 0.9761 & 0.9281  & 0.9272  \\
                                            & $N@20$ & \textbf{0.9775} & 0.9770 & 0.9310  & 0.9297  \\ \hline
\multirow{8}{*}{\begin{tabular}[c]{@{}c@{}}Amazon \\ Automotive\end{tabular}}          & $H@1$    & \textbf{0.9594} & 0.9482 & 0.8933  & 0.8979  \\  
                                            & $H@5$    & \textbf{0.9946} & 0.9903 & 0.9458  & 0.9464  \\ 
                                            & $H@10$   & \textbf{0.9976} & 0.9939 & 0.9515  & 0.9531  \\ 
                                            & $H@20$   & \textbf{0.9989} & 0.9963 & 0.9600  & 0.9618  \\ 
                             \cline{2-6}                & $N@1$  & \textbf{0.9594} & 0.9482 & 0.8933  & 0.8979  \\ 
                                            & $N@5$  & \textbf{0.9897} & 0.9844 & 0.9419  & 0.9422  \\ 
                                            & $N@10$ & \textbf{0.9908} & 0.9857 & 0.9439  & 0.9445  \\ 
                                            & $N@20$ & \textbf{0.9911} & 0.9863 & 0.9460  & 0.9468  \\ \hline
\end{tabular} \caption{Impact of User-Item and Item-Item Instances Self-Attention Mechanisms. $H@K$ and $N@K$ mean $HR@K$ and $NDCG@K$, respectively.}\label{tab:abtest}
\end{table}

\subsection{Variable Sensitivity}
\label{subsec:var_sens}
\begin{figure}[htp]
	\centering
	
	\begin{subfigure}[t]{0.23\textwidth}
		\centering
		\includegraphics[scale=0.24]{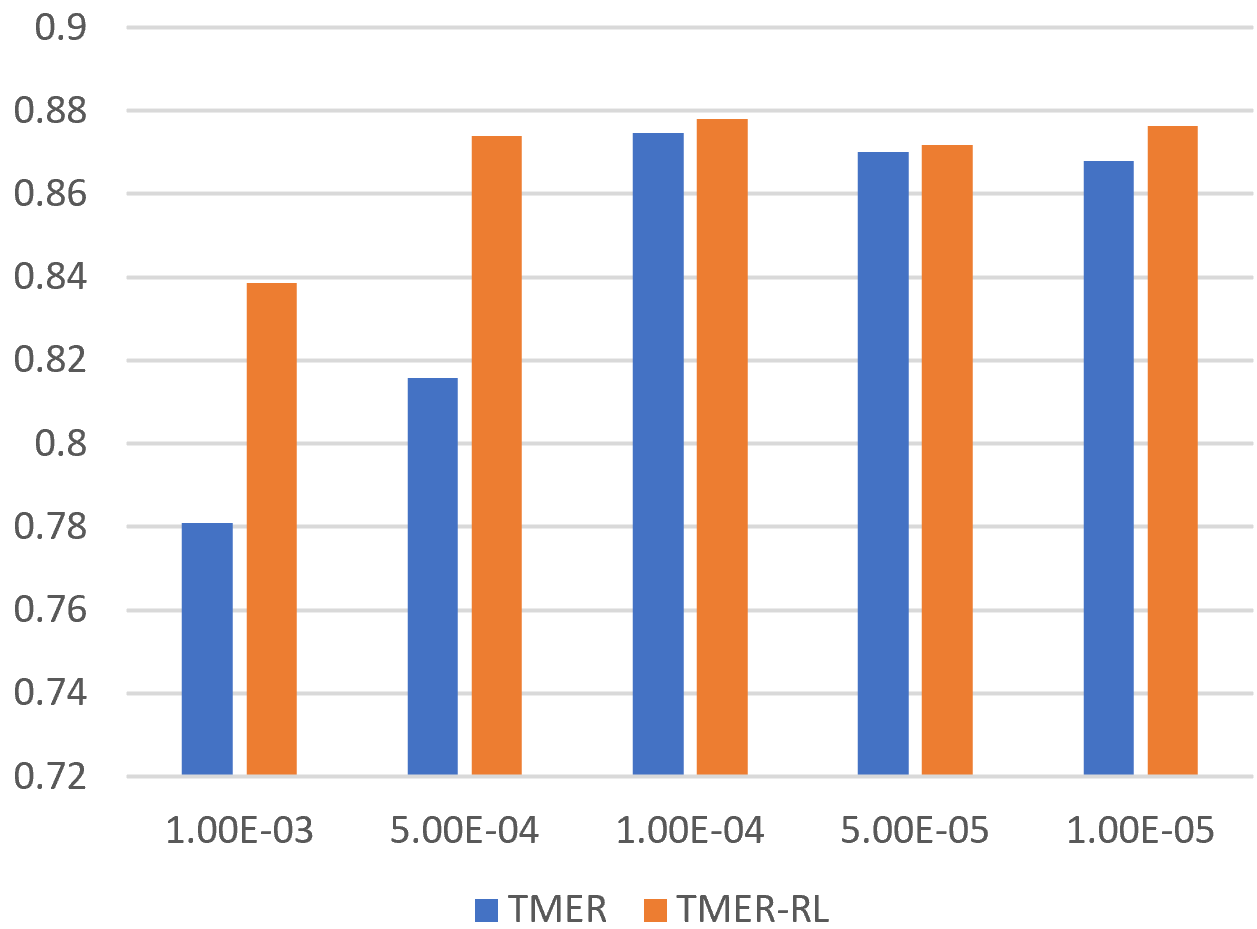} 
		\caption{HR@1 w.r.t. Learning rate on Amazon Musical Instruments} 
		\label{fig:var_sens_exp1-1}
	\end{subfigure}
	~
	\begin{subfigure}[t]{0.23\textwidth}
		\centering
		\includegraphics[scale=0.24]{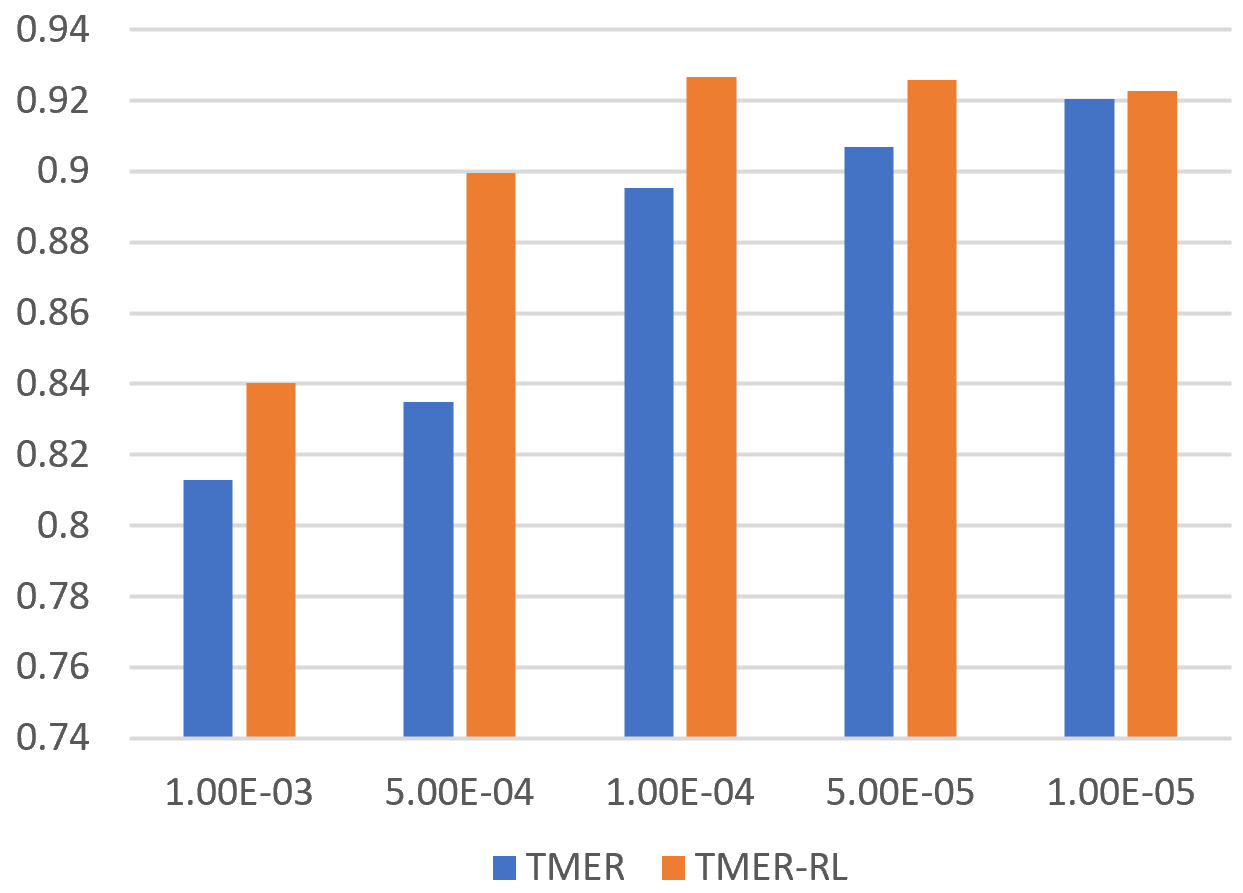} 
		\caption{HR@1 w.r.t. Learning rate on Amazon Automotive} 
		\label{fig:var_sens_exp1-2}
		
	\end{subfigure}
    \caption{HR@1 w.r.t Learning rate} \label{fig:var_sens_exp}
\end{figure}

To test how the recommendation perform when variable changes, we conduct variable sensitivity test on Amazon Musical Instruments dataset and Amazon Automotive dataset. Figure \ref{fig:var_sens_exp} shows the influence of learning rate of recommendation for HR@1. The x-axis represents the learning rate of recommendation, which ranges among \{1e-3, 5e-4, 1e-4, 5e-5, 1e-5\}, and the y-axis means the HR@1 value. 

From Figure \ref{fig:var_sens_exp}, we could see that the HR@1 fluctuation of TMER-RL is smaller than TMER so we could get the conclusion that the performance of TMER-RL is more stable than TMER. Moreover, our proposed TMER-RL could always outperform TMER with different learning rates on these two datasets. This further proves the effectiveness of TMER-RL. Especially, on the larger dataset (Amazon Automotive), TMER-RL is more effective than TMER, which means our reinforcement learning path generation module could show superiority on a larger dataset, which is also required in the real scenario.

\subsection{Impact of the Amounts of Training Data}
\label{subsec:diff_data_proportion}

\begin{figure}[htp]
	\centering
	
	\begin{subfigure}[t]{0.23\textwidth}
		\centering
		\includegraphics[scale=0.24]{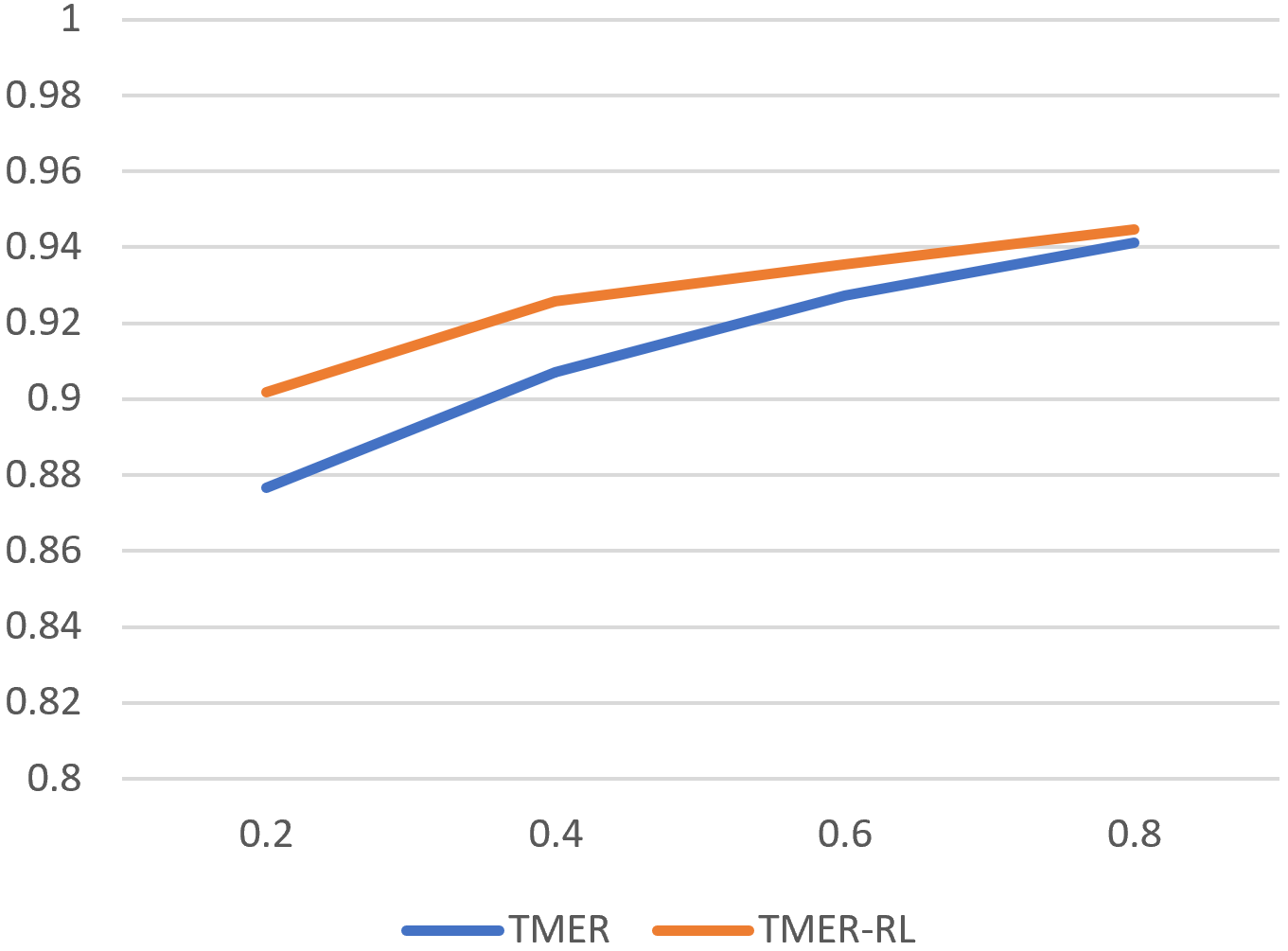} 
		\caption{HR@1 w.r.t. Training Proportion on Amazon Automotive} 
		\label{fig:scalability_automotive_hr1}
	\end{subfigure}
	~
	\begin{subfigure}[t]{0.23\textwidth}
		\centering
		\includegraphics[scale=0.24]{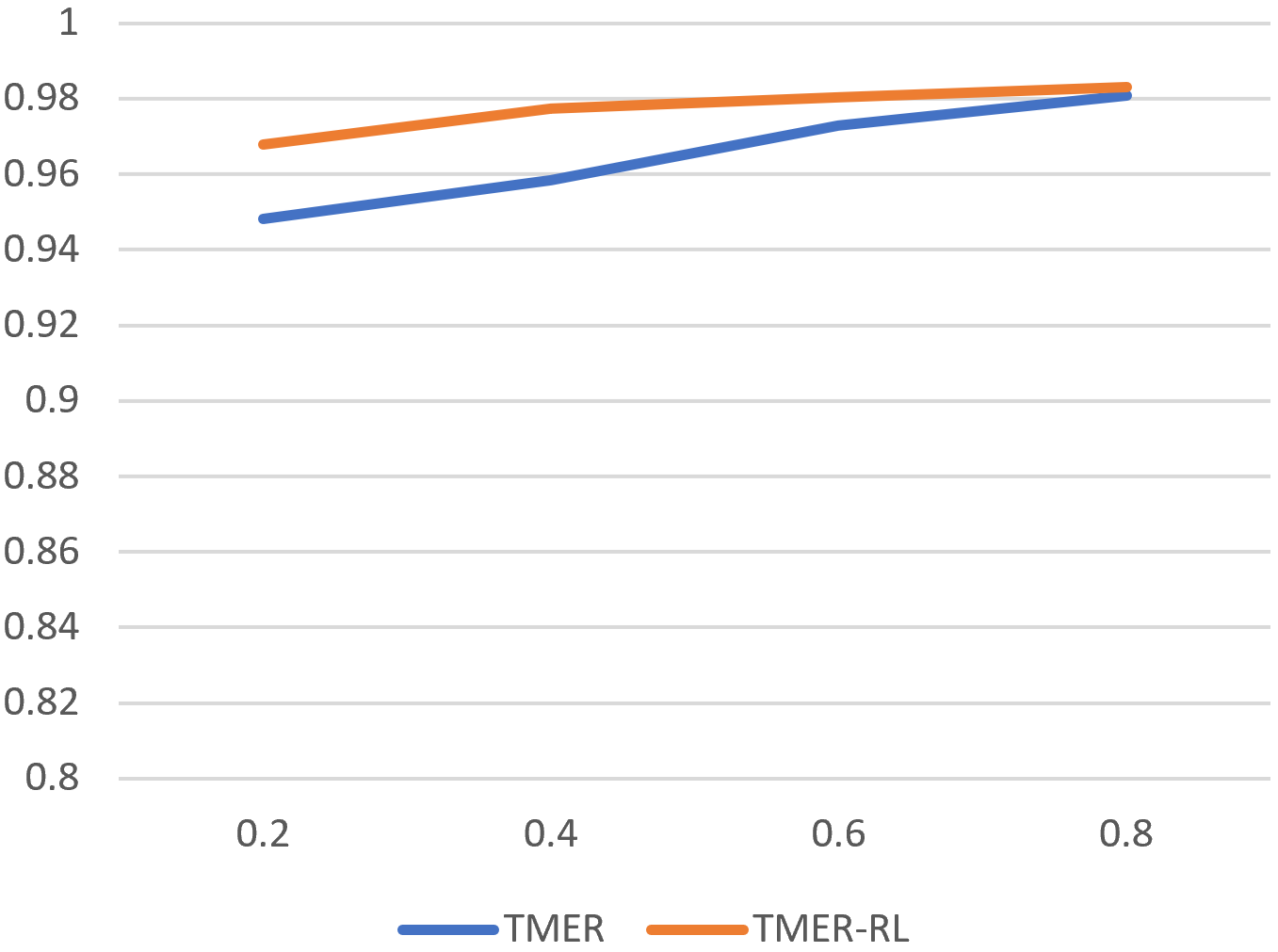} 
		\caption{HR@20 w.r.t. Training Proportion on Amazon Automotive. } 
		\label{fig:scalability_automotive_hr20}
		
	\end{subfigure}
    \caption{HR@1 and HR@20 w.r.t Training Proportion on Amazon Automotive. The x-axis represents the training proportion of the full Amazon Automotive dataset.} \label{fig:scalability_auto}
\end{figure}

To test whether our proposed method TMER-RL could show superiority on sparse dataset, we conduct the TMER-RL and TMER on different training proportion of the full Amazon Automotive dataset. We select \{0.2, 0.4, 0.6, 0.8\} as the training proportion (x-axis in Figure \ref{fig:scalability_auto}) and compare the HR@1 (Figure \ref{fig:scalability_automotive_hr1}) and HR@20 (Figure \ref{fig:scalability_automotive_hr20}) of TMER-RL and TMER.

As we can see, the more training proportion of the full dataset, the better performance of TMER and TMER-RL. This is in line with the common sense because lower training proportion means sparser data, which leads to less paths generation and captures less information among users and items. This will result in lower performance of recommendation. However, in another perspective, when dealing with the least training proportion data (0.2) in Figure \ref{fig:scalability_automotive_hr1} and Figure \ref{fig:scalability_automotive_hr20}, the gap of TMER-RL and TMER is the most and the proposed TMER-RL always keep the better performance. This result could get the result that the TMER-RL is more suitable for sparse dataset, which means it could explore paths in a sparse data environment and do better recommendation. Besides, the TMER-RL could always outperform TMER on HR@1 and HR@20 when the training proportion varies, which further proves the effectiveness of TMER-RL.

\subsection{Case Study: Generating Explanation for Recommendation}

\label{exp:case_study}
One of this work's contributions is to give explanations on instances paths while recommending preferable items. This is because our method generates multiple item-item instance paths according to user purchase history, and then it utilizes attention mechanism to learn the weight of each item-item instance path and aggregate multiple item-item instance paths for each item-item pair. To demonstrate this, we show a random example drawn from TMER on Amazon Musical Instruments dataset.

\begin{figure*}[htp]
	\centering
	\includegraphics[width=0.95\textwidth]{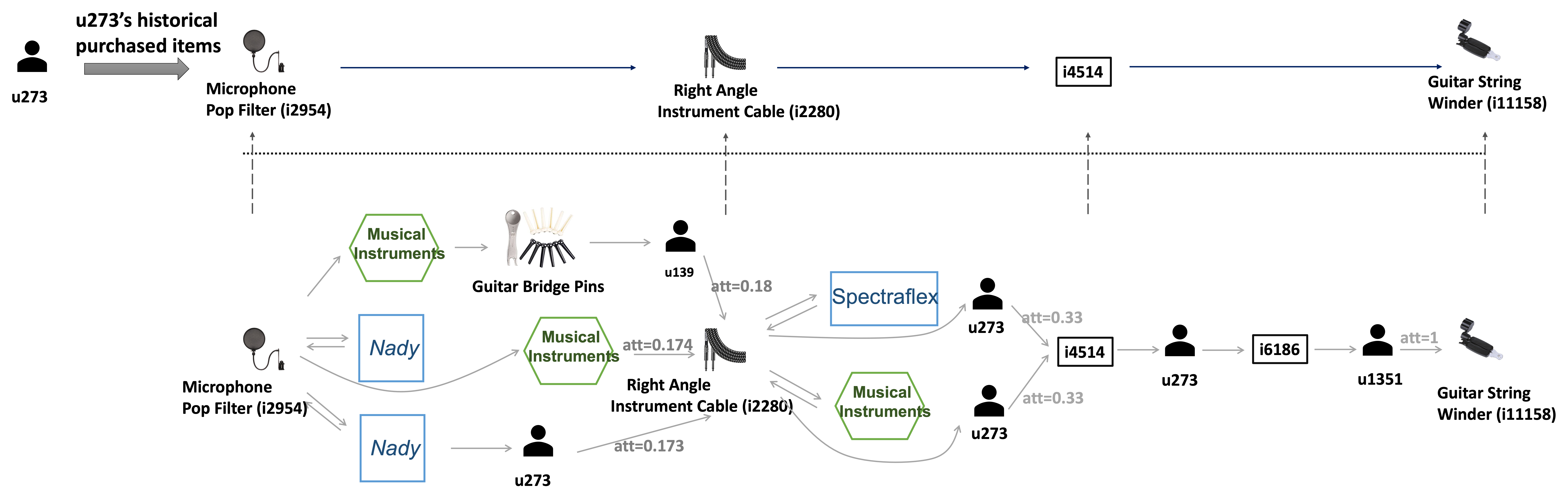}
	\vspace{-1em}
	\caption{It shows $u273$ and related historical purchased items in the upper part of the figure. In the lower part, it displays the reasoning item-item paths along this historical purchased items related to $u273$. Green blocks represent the category attributes. Blue blocks represent the brand attributes. Black blocks without physical pictures do not have meta information in the dataset.} \label{case_study}
	\vspace{-1em}
\end{figure*}

We randomly select a user, whose id is $u273$, and track his item-item instance paths' parameters. In the training dataset, $u273$'s purchase history is $i2954$, $i2280$, $i4514$ and $i11158$. We shows several sampled item-item instance paths with high attention parameters in Figure \ref{case_study} and demonstrates our explanations.

\begin{itemize}
	\item According to the Figure \ref{case_study}, there are three reasons for purchasing $i2280$. The most probable reason with the highest attention weight is that $u273$ has bought the last musical instrument category $i2954$ and $i2237$ with the attention weight $0.18$. For the next item $i4514$, the reason for purchasing it is that the user $u273$ has bought $i2280$ who has the same brand and category with item $i4514$. There is also only one item-item instance path between some items because the item-brand and item-category data are sparse. Therefore, our method can model the reason through item-item instance paths with different weights.
	
	\item Besides, our model can capture sequential information according to user purchase history thanks to item-item instance paths. These item-item instance paths learn the reason path from the current item to the next item. In the whole model, these reason paths will feed to the item attention module. Therefore, our model can recommend with learned sequential information.
\end{itemize}

\subsection{Case Study: Generating Paths via Reinforcement Learning}

\begin{figure}[ht]
	\includegraphics[scale=0.25]{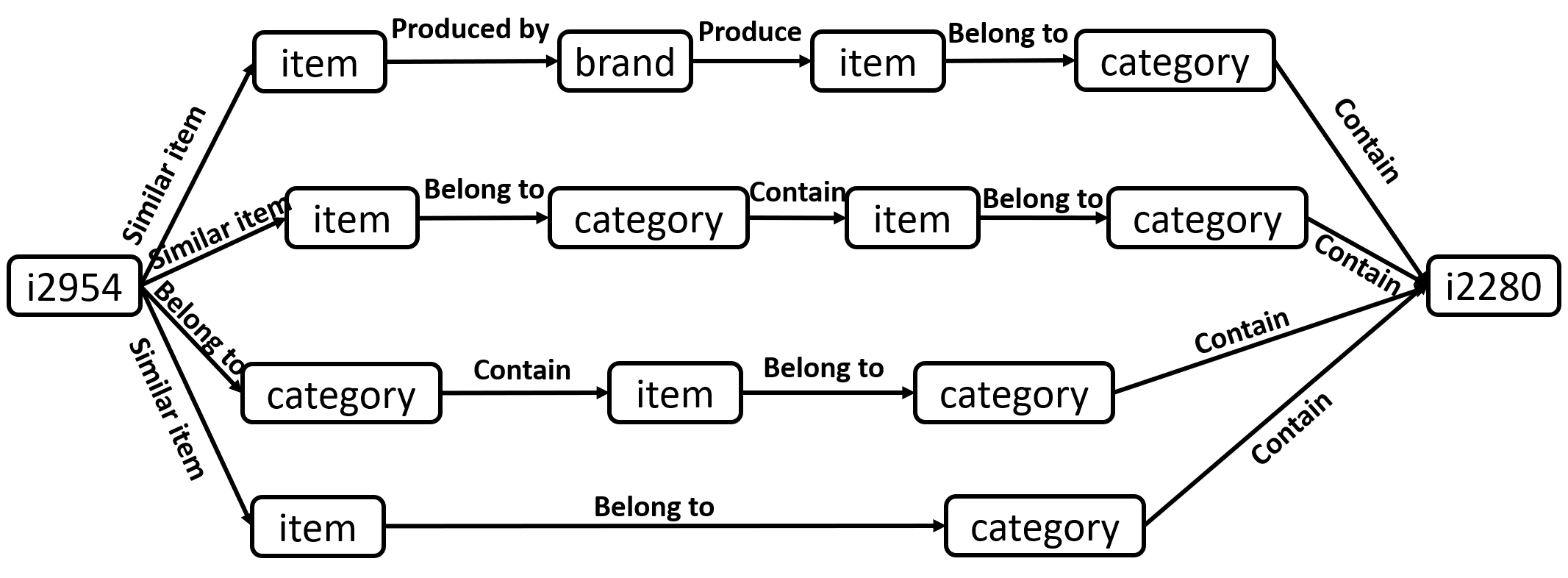}
	\caption{Summary of 10 generated paths via reinforcement learning from i2954 to i2280 for u273.} \label{fig:case_study_rl}
\end{figure}

Compared with randomly generating paths according to pre-defined meta-paths in our previous work \cite{10.1145/3437963.3441762}, we design and implement a reinforcement learning framework to learn and mine instances in this work. Also, instead of randomly choosing paths among large amount of instances, we design a rank method to sort the scores of generated paths and get top $k$ paths to feed into the following modules. We provide a case study for paths generation by reinforcement learning framework on Amazon Musical Instruments dataset. Details are shown in Figure \ref{fig:case_study_rl}. 

To be consistent with \ref{exp:case_study}, we also choose $u273$ as the example. $i2954$ (Microphone Pop filter) and $i2280$ (Right Angle Instrument Cable) are the first and the second items for $u273$. We summarize the generated 10 paths to schemes in Figure \ref{fig:case_study_rl}. The generated paths schemes contain, \textit{IIBICI}, \textit{IICICI}, \textit{ICICI} and \textit{IICI}.

Although the number of summarized schemes is less than that of pre-defined meta-paths, the generated paths via reinforcement learning contributes to recommendation more than randomly generation, which could be proved in the above experiments. To be specific, when randomly generating, the learned paths will be exactly consistent with the pre-defined meta-paths, but it is unsure that the generated paths are useful for the recommendation. However, utilizing reinforcement learning to mine paths could at least guarantee the correlation of nodes on the paths because of the action calculation during exploring. The proposed ranking equation could obtain the $k$ most relevant paths.

Moreover, the pre-defined meta-paths may not be able to define some more useful schemes. For example, in Figure \ref{fig:case_study_rl}, there are 4 schemes, but only \textit{ICICI} is pre-defined meta-path in TMER. Furthermore, if the meta-path defines all existing conditions and considers all situations, the calculating complexity will be very large. Therefore, using reinforcement learning to explore schemes is a better choice.
\section{Related work}
\label{sec_related_work}

Early recommendation systems mostly rely on Collaborative Filtering (CF) \cite{schafer2007collaborative, sarwar2001item}, which are based on the idea that users with similar history will be more likely to purchase similar items. However, CF-based recommendations always have sparsity issues and cold-start problems. Therefore, some works utilize side information, like user and item attributes \cite{gong2009employing}, item contents \cite{melville2002content} to solve this issue. Among them, methods based on knowledge graph \cite{wang2020reinforced, xian2019reinforcement, zhu2020knowledge, wang2019explainable} show great advantages in the recommendation performance and explainability. 

Knowledge graph-based recommendations are roughly be categorized into embedding-based approaches and path-based approaches. Prior efforts on embedding-based knowledge graph recommendations \cite{zhang2016collaborative} always use embeddings of the knowledge graph to model the user-item interactions for recommendations. For example, exiting works \cite{huang2018improving} utilizes TranE \cite{bordes2013translating} to embed user-item interactions to integrate knowledge into the recommendation system. Similarly, \cite{grad2017graph} embeds user and item vectors into the same embedding space for recommendation. The above approaches model the relations of users and items using knowledge graph embedding methods, which achieves great improvement in model expressiveness. However, these methods are sensitive to the quality of related knowledge graphs.

To this end, meta-paths \cite{sun2011pathsim}, and the connectivity of different types of nodes are introduced to recommendations, which have the ability to learn the explicit expression along each meta-path schema. In \cite{zhao2017meta}, the authors introduce Matrix Factorization (MF) and Factorization Machine (FM) to learn similarities generated by each meta-path for recommendation. \cite{han2018aspect} models rich objects and relations in knowledge graph and extracts different aspect-level similarity matrices thanks to meta-paths for the top-N recommendation. Although they achieved appealing performance, the limitation is still obvious that structured meta-path based similarity latent factors can only reflect mutual infectivity along meta-path schemas in a graph but cannot capture the specific information along particular path instances, which limits the explainability of recommendation.

More recently, injecting meta-paths as recommendation context (aggregation of meta-path instances) \cite{hu2018leveraging} was introduced for top-N recommendation. It provides fine-grained explanations based on specific instances. However, it ignores the important sequential dynamics of user-item interactions, which limits the performance of the recommendation performance and interpretability. To consider the sequential information, \cite{wang2019explainable} utilizes LSTM to model the sequential information, but it only considers path-based sequential information between users and items and ignores the importance of the user's clicked history sequences, which are highly informative to infer user's preferences. To tackle this issue, \cite{zhu2020knowledge} attempts to model the sequences of user's behaviours and path connectivity between users and items for recommendation. Nevertheless, it only considers user-item paths, which ignores the item-item intrinsic relation information and cannot learn some complex semantic information between items. Moreover, it has to pre-define some meta-paths and randomly sample meta-path instances, which involves too much human efforts and random factor. Based on the above research, we propose a reinforcement learning based path exploration for recommendation with differentiated user-item and sequential item-item instances to enhance the learning ability for explainability recommendation.

\section{Conclusion}
We propose TMER-RL, which explicitly models dynamic user-item interactions over time with path-based knowledge-aware explainable capabilities. We explore item-item paths between consecutive items with attention mechanisms for users’ sequential behaviour
modelling via a reinforcement learning framework. For evaluation, we conduct 6 sets of experiment to prove the effectiveness of TMER-RL, including comparing with 8 state-of-the-art baselines on 4 datasets to show the high performance, ablation test on modules of TMER-RL to analysis the importance of each module, variable sensitivity test to show the influence of the learning rate varies, impact of the amounts of training data to prove the ability to process sparse data, case study to explain the way to explain recommendation via generated paths and another case study to show the generated paths via reinforcement learning framework. All the above experiments have proved the explainability, effectiveness, high-performance of the TMER-RL. Future works may include the following directions: 1) generate human-readable explanations for recommendation with NLP techniques, and 2) explore causality learning \cite{xu2020causality} to discover
more appealing paths for explainablity.


%



\ifCLASSOPTIONcaptionsoff
  \newpage
\fi



%

\bibliographystyle{ieeetr}
\bibliography{IEEEtran.bib}


%

\begin{IEEEbiography}[{\includegraphics[width=1in,height=1.25in,clip,keepaspectratio]{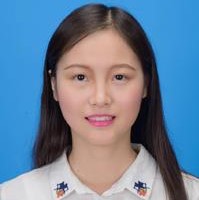}}]{Yicong Li}
is currently a PhD student of Data Science and Machine Intelligence (DSMI) Lab of Advanced Analytics Institute, University of Technology Sydney. She obtained her Master's degree from National University of Defense Technology in 2019. Her research interests mainly focus on data science, graph neural networks, recommender systems, natural language processing and so on. In particular, her current research is focusing on the explainable machine learning, especially the application in recommendation area. She has published papers in international conferences and journals, such as WSDM, KSEM and IEEE Access. She have also reviewed submitted papers in many top-tier conferences and journals, like AAAI, KDD, WWW, IJCAI, WSDM, ICONIP and so on. In addition, she has been invited to review manuscripts in IEEE Transactions on Neural Networks and Learning Systems (TNNLS), which is a top-tier journal in artificial intelligence. 
\end{IEEEbiography}
\vspace{-4em}
\begin{IEEEbiography}[{\includegraphics[width=1in,height=1.25in,clip,keepaspectratio]{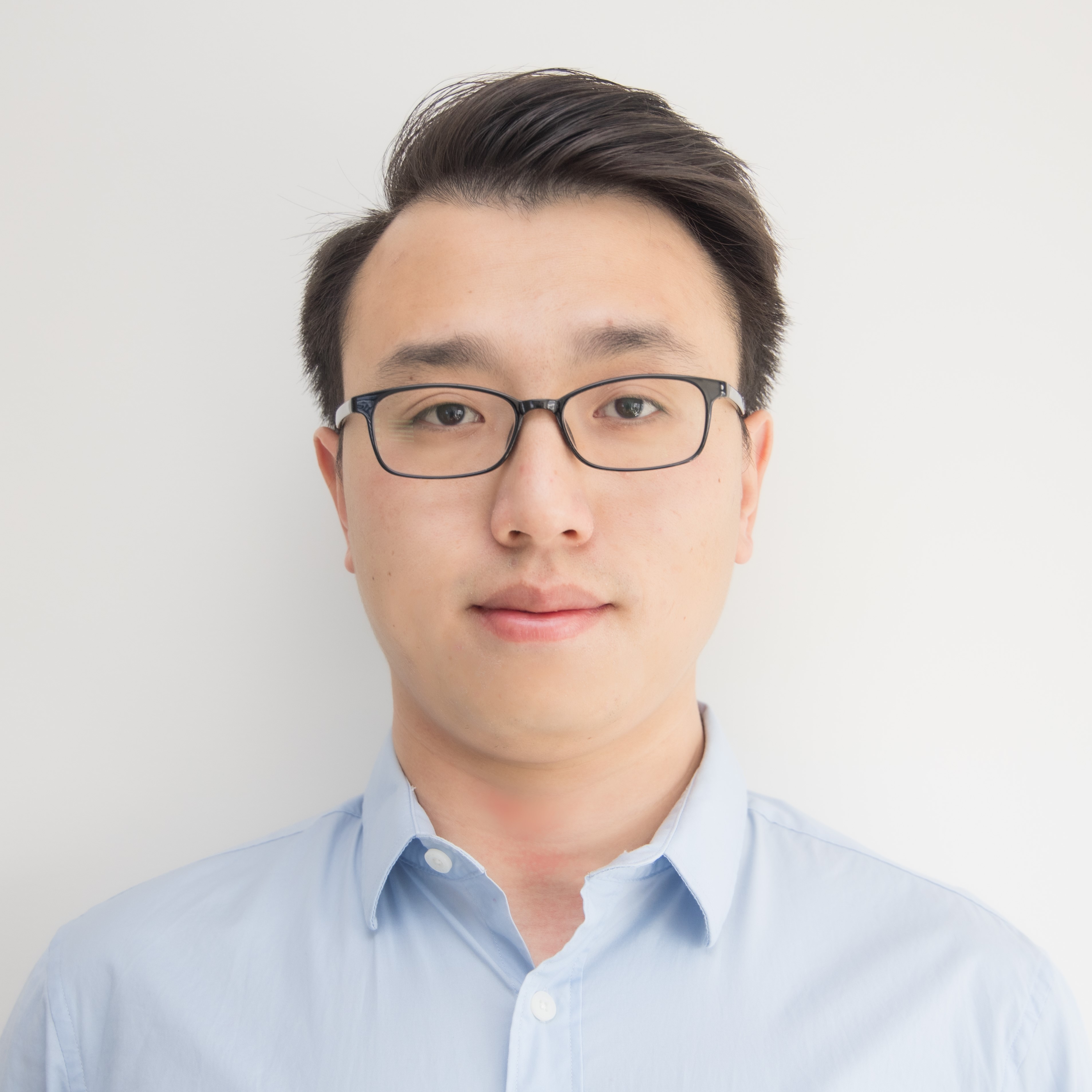}}]{Hongxu Chen}
is a Data Scientist, now working as a Postdoctoral Research Fellow in School of Computer Science at University of Technology Sydney, Australia. He obtained his Ph.D. in Computer Science at The University of Queensland in 2020. His research interests mainly focus on data science in general and expend across multiple practical application scenarios, such as network science, data mining, recommendation systems and social network analytics. In particular, his research is focusing on learning representations for information networks and applying the learned network representations to solve real-world problems in complex networks such as biology, e-commerce and social networks, financial market and recommendations systems with heterogeneous information sources. He has published many peer-reviewed papers in top-tier high-quality international conferences and journals, such as SIGKDD, ICDE, ICDM, AAAI, IJCAI, TKDE. He also serves as program committee member and reviewers in multiple international conference, such as CIKM, ICDM, KDD, SIGIR, AAAI, PAKDD, WISE, and he also acts as invited reviewer for multiple journals in his research fields, including Transactions on Knowledge and Data Engineering (TKDE), WWW Journal, VLDB Journal, IEEE Transactions on Systems, Man and Cybernetics: Systems, Journal of Complexity, ACM Transactions on Data Science, Journal of Computer Science and Technology.
\end{IEEEbiography}
\vspace{-4em}
\begin{IEEEbiography}[{\includegraphics[width=1in,height=1.25in,clip,keepaspectratio]{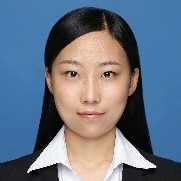}}]{Yile Li}
has received her Master degree from Tongji University, China, in 2021, in transportation engineering, and has received Bachelor's degree from Central South University, China, in 2018. Her research interests include recommendation-system, graph neural network, reinforcement learning and multi-agent system.
\end{IEEEbiography}
\vspace{-4em}
\begin{IEEEbiography}[{\includegraphics[width=1in,height=1.25in,clip,keepaspectratio]{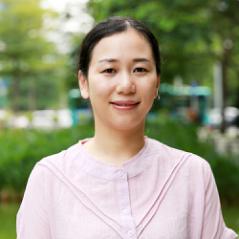}}]{Lin Li}
is a professor in the School of Computer Science and Technology, Wuhan University of Technology, the Distinguished Overseas Researcher of Iwate University, Japan, and the Senior Entrepreneurship Instructor Certified by the Ministry of Human Resources and Social Security. She has received a Ph.D. in Information Science and Technology from the University of Tokyo in 2009.
She has long been engaged in research in the fields of information retrieval, recommendation systems and text mining. She has served as a journal editor for Web Intelligence and Human-Centric Intelligent Systems, and a PC or SPC for international conferences such as IJCAI, AAAI, ACM MM, ICME, COLING, EMNLP. She is a reviewer of ACM TOMM, IPM, ACM TOIS, WWWJ, KBS, etc, and is also a guest editor of WWWJ. She organized a special session for ICME 2021. She has published more than 80 academic papers, and edited or participated in the editing of 2 English monographs (17,094 chapter downloads), 2 Chinese monographs, 4 national invention patents, and more than 10 software copyrights.
\end{IEEEbiography}

\begin{IEEEbiography}[{\includegraphics[width=1in,height=1.25in,clip,keepaspectratio]{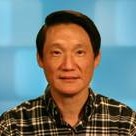}}]{Philip S. Yu}
is a Distinguished Professor in
Computer Science at the University of Illinois at Chicago and also holds the Wexler Chair in Information and Technology. His research interests include data mining, privacy preserving publishing and mining, data streams, database systems, Internet applications and technologies, multimedia systems, parallel and distributed processing, and performance modeling. Dr. Yu has published more than 970 papers in refereed journals and conferences with more than 143,604 citations and an H-index of 1215. He holds or has applied for more than 300 US patents. 
He is a Fellow of the ACM and of the IEEE. He is associate editors of ACM Transactions on the Internet Technology and ACM Transactions on Knowledge Discovery from Data. He is on the steering committee of IEEE Conference on Data Mining and was a member of the IEEE Data Engineering steering committee. He was the Editor-in-Chief of IEEE Transactions on Knowledge and Data Engineering (2001-2004), an editor, advisory board member and also a guest co-editor of the special issue on mining of databases. He had also served as an associate editor of Knowledge and Information Systems.
He also received an IEEE Region 1 Award for "promoting and perpetuating numerous new electrical engineering concepts" in 1999. He had received several UIC honors, including Research of the Year at 2013 and UI Faculty Scholar at 2014. He also received many IBM honors including 2 IBM Outstanding Innovation Awards, an Outstanding Technical Achievement Award, 2 Research Division Awards and the 94th plateau of Invention Achievement Awards. He was an IBM Master Inventor.
Dr. Yu received the B.S. Degree in E.E. from National Taiwan University, the M.S. and Ph.D. degrees in E.E. from Stanford University, and the M.B.A. degree from New York University.
\end{IEEEbiography}
\vspace{-50pt}
\begin{IEEEbiography}[{\includegraphics[width=1in,height=1.25in,clip,keepaspectratio]{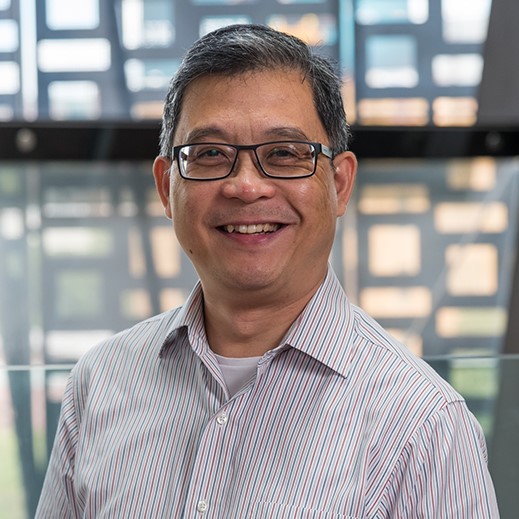}}]{Guandong Xu}
is a professor in the School of Computer Science and Data Science Institute at UTS and an award-winning researcher working in the fields of data mining, machine learning, social computing and other associated fields.
He is Director of the UTS-Providence Smart Future Research Centre, which targets research and innovation in disruptive technology to drive sustainability. 
He also heads the Data Science and Machine Intelligence Lab, which is dedicated to research excellence and industry innovation across academia and industry, aligning with the UTS research priority areas in data science and artificial intelligence.
Guandong has had more than 220 papers published in the fields of Data Science and Data Analytics, Recommender Systems, Text Mining, Predictive Analytics, User Behaviour Modelling, and Social Computing in international journals and conference proceedings in recent years, with increasing citations from academia. 
He has shown strong academic leadership in various professional activities. He is the founding Editor-in-Chief of Human-centric Intelligent System Journal, the Assistant Editor-in-Chief of World Wide Web Journal, as well as the founding Steering Committee Chair of the International Conference of Behavioural and Social Computing Conference.

\end{IEEEbiography}







\end{document}